# HYBRID LBM-FVM AND LBM-MCM METHODS FOR FLUID FLOW AND HEAT TRANSFER SIMULATION

## Zheng Li[a,b], Mo Yang[b] and Yuwen Zhang[a*]

[a] Department of Mechanical and Aerospace Engineering, University of Missouri , Columbia, MO 65211, USA

[b] College of Energy and Power Engineering, University of Shanghai for Science and Technology, Shanghai 200093, China

## ABSTRACT

The fluid flow and heat transfer problems encountered in industry applications span into different scales and there are different numerical methods for different scales problems. It's not possible to use single scale method to solve problems involving multiple scales. Multiscale methods are needed to solve problems involving multiple scales. In this chapter, meso-macro-multiscale methods are developed by combining various single scale numerical methods, including lattice Boltzmann method (LBM), finite volume method (FVM) and Monte Carlo method (MCM). Macroscale methods include FVM, while LBM and MCM belongs to mesoscale methods. Two strategies exist in combing these numerical methods. For the first one, the whole domain is divided into multiple subdomains and different domains use various numerical methods. Message passing among subdomains decides the accuracy of this type of multiscale numerical method. For the second one, various parameters are solved with different numerical methods. These two types of multiscale methods are both discussed in this chapter.

## 1. INTRODUCTION

Development of CFD involves very wide variation of scales [1] ranging from, nano-/micro-, meso- and macroscales. Molecular dynamics (MD) [2] is applicable to nano- and microscale problems and LBM is a typical mesoscopic scale method [3]; Finite difference method (FDM) [4] and FVM [5] on the other hand, falls into the category of macroscale approach [6]. Realistic problems may involve more than one scale and they are so-called multiscale problems. Multiscale transport phenomena exists in many industry areas, such as: fuel cell, laser material interaction and electronics cooling. It is impossible to solve a multiscale problem using any single-scale method. For example, MD simulation cannot be used in the entire simulation domain, and FVM is not suitable for the microscopic region; LBM costs several times more computational time than the FVM to obtain the same accuracy in the macroscopic problem [7]. It's necessary to build multiscale method to solve these multiscale problems.

In this chapter, meso-macro-multiscale methods are developed by combining various single scale numerical methods, including LBM, FVM and MCM while two strategies exist in combing these numerical methods. For the first one, the whole domain is divided into multiple subdomains and different domains use various numerical method. Message passing among subdomains decides the accuracy of this type of multiscale numerical method. For the second one, various parameters are solved with different numerical methods.

For the first strategy of multiscale methods, there are some existing results reported in the literature: MD-FVM [8-10] hybrid methods were developed to combine microscale and macroscale numerical methods. In references [11-13], LBM-MD was proposed to combine mesoscale and microscale numerical methods. This chapter focus on multiscale methods combining mesoscale and macroscale numerical methods. LBM-FDM [14-

---

[*] Corresponding Author. Email: zhangyu@missouri.edu

16] were advanced to solve various multiscale problems. However, the FDM itself has the limitation when solving problems with complex computational domain [6]. This shortfall restricts the development of LBM-FDM because the one of the most attractive advantages of LBM is its suitability to solve the problems in complex computational domain. LBM-FVM [17-19] were developed to solve conduction-radiation heat transfer and compressible fluid flow problems. Luan et al. [20] solved natural convection using the LBM-FVM with the general reconstruction operator [21] reaching persuasive results. Chen et al. [22] used this method to fulfill LBM-FVM with various grids settings. However general reconstruction operator is newly proposed to fulfill the combine method which means more validations are needed for the general reconstruction operator itself. In this chapter, now existing boundary conditions nonequilibrium extrapolation scheme [23] and finite-difference velocity gradient method [16] are employed to fulfill this type of multiscale method.

Second strategy of multiscale method is a new choice for LBM to solve heat transfer problems. This is a promising numerical method because the main advantage of the LBM lies in obtaining a velocity field. Several researchers [24, 25] obtained some good results using the LBM-FDM hybrid method. However, the FDM itself has the limitation when solving problems with complex computational domain [6]. This shortfall restricts the development of LBM-FDM because the one of the most attractive advantages of LBM is its suitability to solve the problems in complex computational domain. There are limited efforts on the hybrid lattice Boltzmann with finite volume method (LBM-FVM) [26, 27] for conduction-radiation problems. Melting problems are solved with LBM-FVM hybrid method [28]. Application of LBM-FVM to natural convection problems will be discussed in this chapter.

Monte Carlo method is a widely used numerical method for heat transfer problem [29]. It can solve conduction, convection and radiation heat transfer problems. In addition, both MCM and LBM belong to mesoscopic scale method. It is possible to combine LBM and MCM to solve convection heat transfer problems, which will also be discussed in this chapter.

## 2. LATTICE BOLTZMANN METHOD

Lattice Boltzmann method is a promising mesoscale method for fluid flow and heat transfer simulation. Instead of solving mass, velocity and energy conservation equations as traditional CFD methods, LBM reaches macroscale parameter using statistical behaviors of particles as shown in Fig. 1, which represent large amounts of fluid molecules.

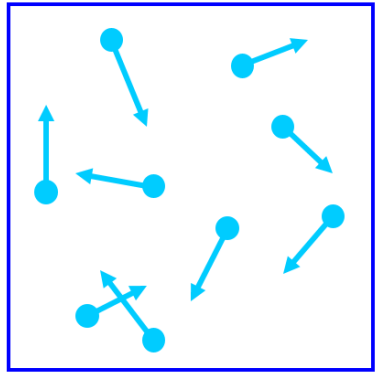

Figure 1 Particle behavior

These particles can steam and collision to each other in the computational domain. It can be employed to solve various fluid flow and heat transfer problems [30-32]. In this section, LBM for fluid flow and heat transfer simulation is included.

## 2.1 Boltzmann equation

Statistical behaviors of particles that are not in thermodynamic equilibrium can be described by the Boltzmann equation:

$$\frac{\partial f}{\partial t} + \boldsymbol{\varepsilon} \cdot \frac{\partial f}{\partial \boldsymbol{r}} + \boldsymbol{a} \cdot \frac{\partial f}{\partial \boldsymbol{\varepsilon}} = \Omega\left(f\right)_{collision} \tag{1}$$

where $f$ is the density distribution, and $\Omega$ is the collision operator that is dictated by the collision rules. $\boldsymbol{\varepsilon}$, $\boldsymbol{r}$ and $\boldsymbol{a}$ are the particle's velocity, location and acceleration, respectively. To solve this Boltzmann equation, we need to simplify the collision term first. In this chapter, we employed the widely used Bhatnagar-Gross-Krook model to fulfill this process.

## 2.2 Lattice Bhatnagar-Gross-Krook model

The Bhatnagar-Gross-Krook model (BGK) that uses the Maxwell equilibrium distribution, $f^{eq}$, will be used in this chapter:

$$f^{eq} = n \frac{1}{\left(2\pi R_g T\right)^{m/2}} \exp\left[-\frac{\left(\boldsymbol{\varepsilon} - \boldsymbol{u}\right)^2}{2R_g T}\right] \tag{2}$$

where m is the dimension of the problem. Equation (2) describes the situation that the system has reached to the final equilibrium. The BGK model assumes that the collision term is the time relaxation from density distribution to the Maxwell equilibrium distribution. Assuming the relaxation time is $\tau_v$, the Boltzmann equation under the BGK model can be expressed as:

$$\frac{\partial f}{\partial t} + \boldsymbol{\varepsilon} \cdot \frac{\partial f}{\partial \boldsymbol{r}} + \boldsymbol{a} \cdot \frac{\partial f}{\partial \boldsymbol{\varepsilon}} = -\frac{1}{\tau_v}\left(f - f^{eq}\right) \tag{3}$$

The LBM used in this chapter is a special scheme of LBGK. Only limited numbers of directional derivatives are applied to Eq. (3) and there must be enough information to obtain the macroscopic governing equation.

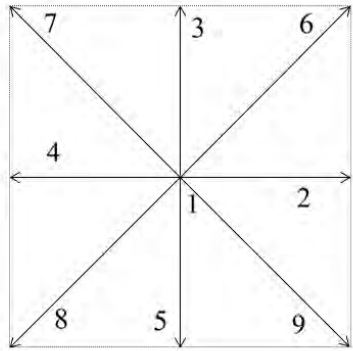

Figure 2 Nine directions in D2Q9 model

The D2Q9 model is used and nine directions are selected in the 2-D problem shown in the Fig. 2. The velocity in every direction is:

$$e_i = \begin{cases} (0,0) & i = 1 \\ c(-\cos\dfrac{a\pi}{2}, -\sin\dfrac{a\pi}{2}) & i = 2,3,4,5 \\ \sqrt{2}c(-\cos\dfrac{(2a+1)\pi}{4}, -\sin\dfrac{(2a+1)\pi}{4}) & i = 6,7,8,9 \end{cases} \tag{4}$$

where $c$ is a constant in the lattice unit. The density distribution $f_i$ in the fixed direction can be obtained by integrating Eq. (3):

$$f_i(r + e_i\Delta t, t + \Delta t) - f_i(r,t) = -\frac{1}{\tau_v}\Big(f_i^{eq}(r,t) - f_i(r,t)\Big) + \Delta t F_i(r,t) \tag{5}$$

where $\Delta t$ is the magnitude of the time step while $F_i$ is the body force in the fixed direction. When the velocity is low, $f_i^{eq}$ can be simplified as:

$$f_i^{eq} = \rho\omega_i\left[1 + \frac{e_i \cdot u}{R_g T} + \frac{(e_i \cdot u)^2}{2R_g^2 T^2} - \frac{u^2}{R_g T}\right] \tag{6}$$

$$\omega_i = \left(2\pi R_g T\right)^{-\frac{m}{2}} \exp\left(-\frac{e_i^2}{2R_g T}\right) \tag{7}$$

where $f_i^{eq}$ can be further simplified with regarding $c_s$ that is the speed of sound in the lattice unit:

$$f_i^{eq} = \rho\omega_i\left[1 + \frac{e_i \cdot u}{c_s^2} + \frac{(e_i \cdot u)^2}{2c_s^4} - \frac{u^2}{2c_s^2}\right] \tag{8}$$

where

$$\omega_i = \begin{cases} \dfrac{4}{9} & i = 1 \\ \dfrac{1}{9} & i = 2,3,4,5 \\ \dfrac{1}{36} & i = 6,7,8,9 \end{cases} \tag{9}$$

## 2.3 Chapman-Enskog Expansion

Applying the following Chapman-Enskog expansion equations

$$\frac{\partial}{\partial r} = K\frac{\partial}{\partial r_1} \tag{10}$$

$$\frac{\partial}{\partial t} = K\frac{\partial}{\partial t_1} + K^2\frac{\partial}{\partial t_2} \tag{11}$$

$$f_a = f_a^0 + K f_a^1 + K^2 f_a^2 \tag{12}$$

to Eq. (5), the macroscopic governing equations can be obtained from LBM:

$$\frac{\partial \rho}{\partial t} + \nabla \cdot (\rho \boldsymbol{V}) = 0 \tag{13}$$

$$\frac{\partial (\rho \boldsymbol{V})}{\partial t} + \nabla \cdot (\rho \boldsymbol{V} \boldsymbol{V}) = -\nabla p + \nabla \cdot \left[ \rho \nu \left( \nabla \boldsymbol{V} + (\nabla \boldsymbol{V})^T - \frac{\nu}{c_s^2} \nabla \cdot (\rho \boldsymbol{V} \boldsymbol{V} \boldsymbol{V}) \right) \right] \tag{14}$$

$$\nu = c_s^2 \left( \tau_\nu - \frac{1}{2} \right) \Delta t \tag{15}$$

Equation (14) differs from the macroscopic momentum conservation from due to presence of the term $\nabla \cdot \left[ \rho \nu \left( -\frac{\nu}{c_s^2} \nabla \cdot (\rho \boldsymbol{V} \boldsymbol{V} \boldsymbol{V}) \right) \right]$. Fortunately, it can be neglected when Mach number is low, which is the case in consideration. Thus, the LBM satisfies the same macroscopic governing equations as the macroscopic method, which meets the requirement to combine LBM with FVM.

To obtain the macroscopic parameter, the following two additional equations are needed.

$$\rho = \sum_{i=1}^{9} f_i \tag{16}$$

$$\rho \mathbf{V} = \sum_{i=1}^{9} e_i f_i \tag{17}$$

where $e_i$ is the vector representing the velocity in every discrete direction and $\mathbf{V}$ is the vector of the macroscopic velocity.

## 2.4 Thermal LBM model

Two distribution functions are selected for the fluid flow and heat transfer in LBM. The density and energy distributions are represented by $f_i$ and $g_i$, which are related by the buoyancy force. D2Q9 model for $f_i$ has been included in Section 2.1.

The buoyancy force can be obtained as:

$$F_i = \Delta t \boldsymbol{G} \cdot \frac{(\boldsymbol{e}_i - \boldsymbol{V})}{p} f_i^{eq} \tag{18}$$

where the pressure, $p$, equals $\rho c_s^2$.

The D2Q5 model [33] is used for the temperature field. There are five discrete velocity at each computing node shown is Fig. 2-3.

Similar to the density distribution, the energy distribution can be obtained by

$$g_i(\boldsymbol{r} + \boldsymbol{e}_i \Delta t, t + \Delta t) - g_i(\boldsymbol{r}, t) = \frac{1}{\tau_T} \left( g_i^{eq}(\boldsymbol{r}, t) - g_i(\boldsymbol{r}, t) \right), \quad i = 1, 2, \dots 5 \tag{19}$$

The macroscopic energy equation can be obtained from by Eq. (19) using Chapman-Enskog expansion.

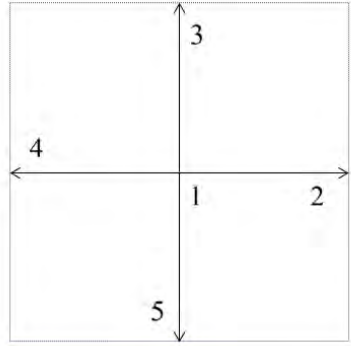

Figure 3 Five directions in D2Q5 model

Then the relaxation time $\tau_T$ is related to the thermal diffusivity $\alpha$ as

$$\alpha = c_s^2\left(\tau_T - \frac{1}{2}\right)\Delta t \qquad (20)$$

The equilibrium energy distribution in Eq. (19) is

$$g_i^{eq} = T\omega_i^T\left(1 + \frac{\boldsymbol{e}_i \cdot \boldsymbol{V}}{c_s^2}\right) \qquad (21)$$

where:

$$\omega_i^T = \begin{cases} \dfrac{1}{3} & i = 1 \\[2mm] \dfrac{1}{6} & i = 2,3,4,5 \end{cases} \qquad (22)$$

The temperature at each computing node can be obtained as:

$$T = \sum_{i=1}^{5} g_i \qquad (23)$$

Two dimensional double distributions LBM model for fluid flow and heat transfer simulation is introduced in this Section 2.

## 3. TWO SCHEMES FOR HYBRID LATTICE BOLTZMANN and FINITE VOLUME METHODS

Two schemes for hybrid LBM-FVM method are proposed in this Section 3. The key point of the hybrid method is to pass the information on the interface between LBM and FVM. It is difficult to transfer velocity obtained from FVM into node population that is needed in LBM. Nonequilibrium extrapolation scheme [23] and finite-difference velocity gradient method [16] are boundary conditions for LBM. They will be used in this section to pass information between the FVM and LBM zones. The lid-driven flow problem is solved to test the proposed methods.

## 3.1 Finite volume method

Lattice Boltzmann method for 2-D fluid flow has been included in Section 2. SIMPLE is a very popular FVM algorithm [34] that solves the general equations in macroscopic scale based on the control volume shown in Fig. 4. This algorithm is employed in this Chapter as FVM.

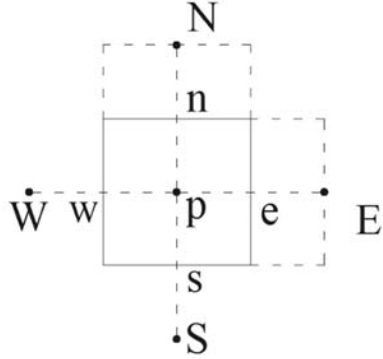

Figure 4 Control volume in 2-D FVM

For a 2-D problem in Cartesian coordinate system, the general equations can be expressed as:

$$\frac{\partial \Phi}{\partial t} + \frac{\partial (\rho u \Phi)}{\partial x} + \frac{\partial (\rho v \Phi)}{\partial y} = \frac{\partial}{\partial x}\left(\Gamma \frac{\partial \Phi}{\partial x}\right) + \frac{\partial}{\partial y}\left(\Gamma \frac{\partial \Phi}{\partial y}\right) + S \tag{24}$$

The SIMPLE algorithm is employed to solve Eq. (24) in this chapter. QUICK scheme [6] is selected to have an accuracy good enough.

## 3. 2 Combined LBM and FVM method for pure fluid flow problems

### 3.2.1 Problem Statement

Lid driven flow is used to test the method to combine LBM and FVM. No-slip boundary conditions are applied to this 2-D problem, and the flow is driven by a constant lid velocity $u_0$ on the top of the square cavity while the velocities on all other boundaries are zero shown in Fig. 5.

This problem can be described by the following governing equations:

$$\frac{\partial \rho}{\partial t} + \frac{\partial (\rho u)}{\partial x} + \frac{\partial (\rho v)}{\partial y} = 0 \tag{25}$$

$$\frac{\partial (\rho u)}{\partial t} + \frac{\partial (\rho u u)}{\partial x} + \frac{\partial (\rho v u)}{\partial y} = -\frac{\partial p}{\partial x} + \mu \left(\frac{\partial^2 u}{\partial x^2} + \frac{\partial^2 u}{\partial y^2}\right) \tag{26}$$

$$\frac{\partial (\rho v)}{\partial t} + \frac{\partial (\rho u v)}{\partial x} + \frac{\partial (\rho v v)}{\partial y} = -\frac{\partial p}{\partial y} + \mu \left(\frac{\partial^2 v}{\partial x^2} + \frac{\partial^2 v}{\partial y^2}\right) \tag{27}$$

which are subject to the following boundary conditions:

$$x = 0 \quad u = 0 \quad v = 0 \tag{28}$$

$$x = H \quad u = 0 \quad v = 0 \tag{29}$$

$$y = 0 \quad u = 0 \quad v = 0 \tag{30}$$

$$y = H \quad u = u_0 \quad v = 0 \tag{31}$$

In addition, the Reynolds number is defined with the constant velocity on the top.

$$\text{Re} = \frac{u_0 h}{v} \tag{32}$$

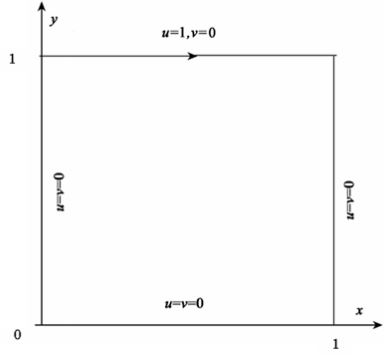

Figure 5 Lid-driven flow

### 3.2.2 Description of the combing method

Lattice unit is applied to LBM while SIMPLE algorithm uses non-dimensional procedure. As discussed above, $e_i$ has different values in different directions in the lattice unit. In order to combine these two methods with different units together, they must be used to describe the same situation in the actual unit.

In the LBM unit conversion process, which changes all the properties into lattice unit, the speed of sound $c_s$ and time step $\Delta t$ are fixed so that the density distributions are all on the computational nodes.

$$c_s = 1/\sqrt{3}, \quad \Delta t = 1 \tag{33}$$

Assuming that the real speed of sound is $u_{sound}$ and the number of nodes in the $y$-direction is $n+1$, the dimensionless velocities in LBM are:

$$U_L = \frac{u}{\sqrt{3}u_{sound}} \tag{34}$$

$$V_L = \frac{v}{\sqrt{3}u_{sound}} \tag{35}$$

and the time step is

$$\Delta t = \frac{\dfrac{H_L}{n}}{\sqrt{3}c_s} = 1 \tag{36}$$

Thus the coordinates in lattice unit become

$$X_L = \frac{nx}{H} \tag{37}$$

$$Y_L = \frac{nx}{H} \tag{38}$$

In order to allow the boundary velocity in lattice unit in LBM to be same as the dimensionless velocity in SIMPLE, it is assuming that the lid velocity and the speed of sound have the following relationship:

$$10u_0 = \sqrt{3}u_{sound} \tag{39}$$

which requires that $10u_0$ to be used in the non-dimension process. The dimensionless lid velocity in the LBM becomes

$$U_0 = 0.1 \tag{40}$$

At this point, the only unknown parameter in lattice unit is the kinematic viscosity $\nu_L$ and it can be obtained from the Reynolds number:

$$\nu_L = \frac{U_0 H_L}{\mathrm{Re}_L} \tag{41}$$

To satisfy Eq. (15), the relaxation time $\tau_\nu$ can be obtained to fulfill LBM

$$\tau_0 = 3\nu_L + 0.5 \tag{42}$$

As discussed above, to obtain the same boundary dimensionless velocity as that in lattice unit, the following non-dimensional variables are defined:

$$\left. \begin{array}{l} U_s = \dfrac{u}{10u_0}, \ V_s = \dfrac{v}{10u_0}, \ X_s = \dfrac{x}{H}, \ Y_s = \dfrac{y}{H} \\[2mm] F = \dfrac{t}{H/(10u_0)}, \ P = \dfrac{p}{\rho(10u_0)^2}, \ \mathrm{Re}_s = \dfrac{u_0 H}{\nu} \end{array} \right\} \tag{43}$$

The governing equations (24) to (30) can be nondimenionalized as:

$$\frac{\partial U_s}{\partial X_s} + \frac{\partial V_s}{\partial Y_s} = 0 \tag{44}$$

$$\frac{\partial U_s}{\partial F} + \frac{\partial \left(U_s U_s\right)}{\partial X_s} + \frac{\partial \left(V_s U_s\right)}{\partial Y_s} = -\frac{\partial P}{\partial X_s} + \frac{1}{10\mathrm{Re}_s}\left(\frac{\partial^2 U_s}{\partial X_s^2} + \frac{\partial^2 U_s}{\partial Y_s^2}\right) \tag{45}$$

$$\frac{\partial V_s}{\partial F} + \frac{\partial \left(U_s V_s\right)}{\partial X_s} + \frac{\partial \left(V_s V_s\right)}{\partial Y_s} = -\frac{\partial P}{\partial Y_s} + \frac{1}{10\mathrm{Re}_s}\left(\frac{\partial^2 V_s}{\partial X_s^2} + \frac{\partial^2 V_s}{\partial Y_s^2}\right) \tag{46}$$

$$X_s = 0 \ \ U_s = 0 \ \ V_s = 0 \tag{47}$$

$$X_s = 1 \ \ U_s = 0 \ \ V_s = 0 \tag{48}$$

$$Y_s = 0 \ \ U_s = 0 \ \ V_s = 0 \tag{49}$$

$$Y_s = 1 \quad U_s = 0.1 \quad V_s = 0 \tag{50}$$

In order to meet the requirement for describing the same situation in the actual unit in LBM and SIMPLE, the following two additional equations are needed.

$$\mathrm{Re}_s = \mathrm{Re}_L \tag{51}$$

$$\Delta F = \frac{\Delta t}{n} \tag{52}$$

where $n+1$ is the number of nodes in the $y$-direction. By following the above nondimensionalization procedures for LBM and SIMPLE, the non-dimensional lid velocity in both methods are 0.1. So the same non-dimensional velocities are reached at the same real location for the same Reynolds number in every time step. Therefore, the non-dimensional velocities can be transferred directly.

In order to combine LBM and SIMPLE in the same problem, the computational domain is divided into two zones as shown in Fig. 6.

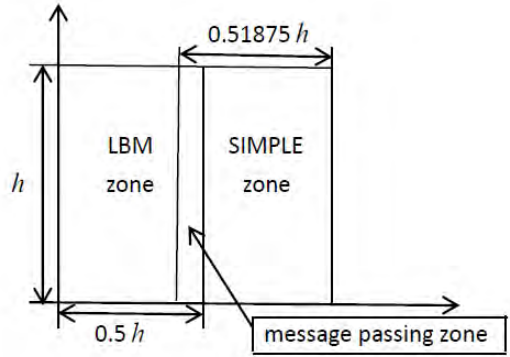

Figure 6 Computational domain for LBM and SIMPLE

A $160 \times 160$ uniform grid is applied to the entire computational domain. In most cases, the wider the message passing zone is, the better the accuracy is. However, enlarged message passing zone also increases the computational time in every time step. Meanwhile, although the grid is uniform in the whole domain, the locations of velocity on the grid are different for LBM and SIMPLE. So additional interpolation, which may require the information on the nearby nodes, is needed in the information sharing process. Three shared grids are selected after testing and the grid in LBM is $80 \times 160$ while that in SIMPLE is $83 \times 160$. It is necessary to point out that it is almost impossible to transfer the pressure in LBM to that in SIMPLE because they have different ways to obtain the pressure. In LBM, the ideal gas law is used to obtain the pressure, while SIMPLE solves the pressure correction equation based on the conservation of mass. The example in this section does not need to transfer the pressure information in the message passing zone shown in Fig. 7 due to the nature of method used to combine LBM and SIMPLE. The interface between LBM and SIMPLE zone is treated as a fixed velocity boundary at every time step. Staggered grids are used in SIMPLE and LBM, and the locations of macroscopic parameters in the computational domain are shown in Figs. 7 and 8. It can be seen that the locations of macroscopic parameters are different in SIMPLE and LBM, even the grids are the same. Figure 9 gives a more clear view about that in one control volume in the message passing zone.

The boundaries of the LBM and SIMPLE zones are $l_1$ and $l_2$ as shown in Fig. 6. Meanwhile, they are the inner nodes in both SIMPLE and LBM zones. In addition, it can also be seen that there is no difference between boundaries and inner nodes for LBM as can be seen from Fig. 8. Therefore, transferring information from SIMPLE to LBM only needs the information on $I_1$ from the nearby nodes in the SIMPLE zone:

$$u_{i,j,L} = \frac{\left(u_{i+1,j+1,S} + u_{i+1,j,S}\right)}{2} \tag{53}$$

$$v_{i,j,L} = \frac{\left(v_{i+1,j+1,S} + v_{i,j+1,S}\right)}{2} \tag{54}$$

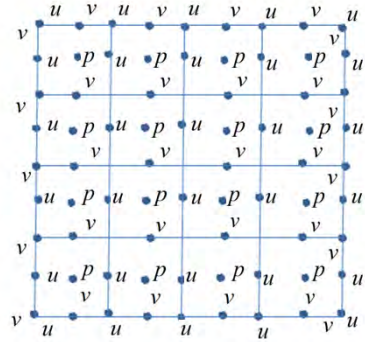

Figure 7 Variable locations in SIMPLE

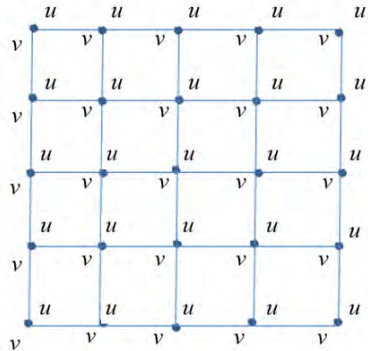

Figure 8 Computational nodes in LBM

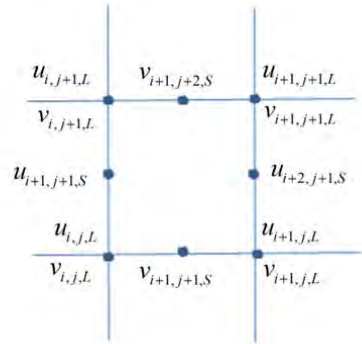

Figure 9 Details on one control volume in the message passing zone

The procedure to transfer information from LBM to SIMPLE is similar except the locations of nodes are different for inner nodes and boundary nodes as shown in Fig. 6. The following equations can be used in this transfer process:

$$u_{i,j,S} = \frac{\left(u_{i-1,j-1,L} + u_{i-1,j,L}\right)}{2} \tag{55}$$

$$v_{i,j,S} = v_{i,j-1,L} \tag{56}$$

After transferring the information between the two zones, the hybrid method turns to fixed velocity problem on the interfaces $l_1$ and $l_2$ shown in Fig. 6 in LBM and SIMPLE. There is no need for any special treatment to the SIMPLE zone rather than setting the boundary velocity in the program. On the other hand, it is different for the LBM zone because its original variable is density distribution on the computational nodes.

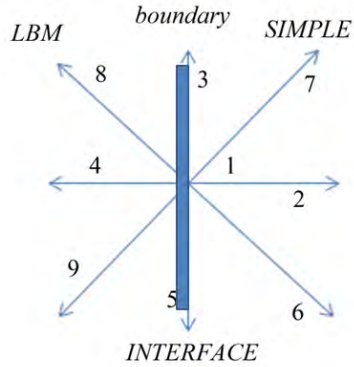

Figure 10 Boundary condition in LBM at the interface

The three density distributions $f_4$, $f_8$, $f_9$ and density $\rho$ are unknown as shown in Fig. 10, while there are only three equations:

$$f_4 + f_8 + f_9 = \rho - f_1 - f_2 - f_3 - f_5 - f_6 - f_7 \tag{57}$$

$$f_4 + f_8 + f_9 = -\rho u_L + f_2 + f_6 + f_7 \tag{58}$$

$$f_8 - f_9 = \rho v_L + f_5 + f_6 - f_3 - f_7 \tag{59}$$

There are several methods to solve this problem on the boundary [16]. Nonequilibrium extrapolation scheme and finite-difference velocity gradient method are used in this section to replace all the density distributions on the boundary.

In the Nonequilibrium extrapolation scheme (hybrid method 1), we have the following assumption:

$$\left(f_i - f_i^{eq}\right)_{boundary} = \left(f_i - f_i^{eq}\right)_{inner} \quad i = 1, 2 \cdots 9 \tag{60}$$

This scheme has second order accuracy and can meet the requirement of the combining method.

One the other hand, finite-difference velocity gradient method (hybrid method 2) is used to replace all the density distributions on the boundary.

Equation (12) can be rewritten as:

$$f_i = f_i^{(0)} + K f_i^{(1)} + O(K^2) \tag{61}$$

The following equations can be obtained depending on different order of $K$.

$$K^0: \ f_i^{(0)} = f^{eq} \tag{62}$$

$$K^1: \ \left( \frac{\partial}{\partial t_1} + \boldsymbol{e}_i \cdot \nabla_1 \right) f^{eq} + \frac{1}{\tau \Delta t} f^{(1)} = 0 \tag{63}$$

$$K^2: \ \frac{\partial f^{eq}}{\partial t_2} + \left( \frac{\partial}{\partial t_1} + \boldsymbol{e}_i \cdot \nabla_1 \right) \left( 1 - \frac{1}{2\tau} \right) f^{(1)} + \frac{1}{\tau \Delta t} f^{(2)} = 0 \tag{64}$$

Then $f^{(1)}$ can be obtained from Eq. (63):

$$f_i^{(1)} = -\tau \Delta t \left( \frac{\partial}{\partial t_1} + \boldsymbol{e}_a \cdot \nabla_1 \right) f_i^{eq} \tag{65}$$

The hydrodynamic variables also have the relation with density distributions:

$$\boldsymbol{\Pi} = \sum_{I=1}^{9} \boldsymbol{e}_i \boldsymbol{e}_i f_i \tag{66}$$

where $\boldsymbol{\Pi}$ is the moment of order 2.

Assume that:

$$\boldsymbol{Q}_i = \boldsymbol{e}_i \boldsymbol{e}_i - c_s^2 \boldsymbol{I} \tag{67}$$

where $\boldsymbol{I}$ is the identity tensor,

one obtains:

$$K f_i^{(1)} = -\frac{\tau \omega_i}{c_s^2} \left( \boldsymbol{Q}_i : \rho \nabla \boldsymbol{u} - \boldsymbol{e}_i \nabla : \rho \boldsymbol{u} \boldsymbol{u} + \frac{1}{2c_s^2} (\boldsymbol{e}_i \cdot \nabla)(\boldsymbol{Q}_i : \rho \boldsymbol{u} \boldsymbol{u}) \right) \tag{68}$$

Substituting Eq. (62) to Eqs. (17) and (66) and approximating $K f_i^{(1)}$ by the first term only [16], the density distribution can be expressed as:

$$f_i = f_i^{eq} - \frac{\tau \omega_i}{c_s^2} \boldsymbol{Q}_i : \rho \nabla \boldsymbol{u} \tag{69}$$

So the density distribution can be related to strain rate tensor $\boldsymbol{S}$ due to the symmetry of $\boldsymbol{Q}_i$:

$$f_i = f_i^{eq} - \frac{\tau \omega_i}{c_s^2} \boldsymbol{Q}_i : \boldsymbol{S} \tag{70}$$

where

$$\boldsymbol{S} = \frac{1}{2} \left( \nabla \boldsymbol{u} + (\nabla \boldsymbol{u})^T \right) \tag{71}$$

Thus, the information of the strain rate tensor on $l_2$ is needed to be transferred to LBM zone from the FVM zone.

In addition, more information needs to be transferred from the SIMPLE zone to LBM zone besides the velocity. $\dfrac{\partial u}{\partial x}$, $\dfrac{\partial u}{\partial y}$, $\dfrac{\partial v}{\partial x}$ and $\dfrac{\partial v}{\partial y}$ on $l_1$ are also needed in order to get the strain rate tensor. $\dfrac{\partial u}{\partial y}$, $\dfrac{\partial v}{\partial x}$ and $\dfrac{\partial u}{\partial x}$ be approximated by a centered difference. The conservation of mass requires that:

$$\frac{\partial v}{\partial y} = -\frac{\partial u}{\partial x} \qquad (72)$$

The following steps should be taken to transfer information between the two zones in the method to combine LBM and SIMPLE (see Fig. 6):

1.  Assume the velocity on $x = l_2$.
2.  Use SIMPLE to solve the velocity in the SIMPLE zone.
3.  Transfer the information on $x = l_1$ from the SIMPLE zone to the LBM zone.
4.  Use the LBM to solve the velocity in the LBM zone.
5.  Transfer the information on $l_2$ from LBM zone to SIMPLE zone.
6.  Go back to step 2 until the velocity is converged.

### 3.3 Two schemes results comparison

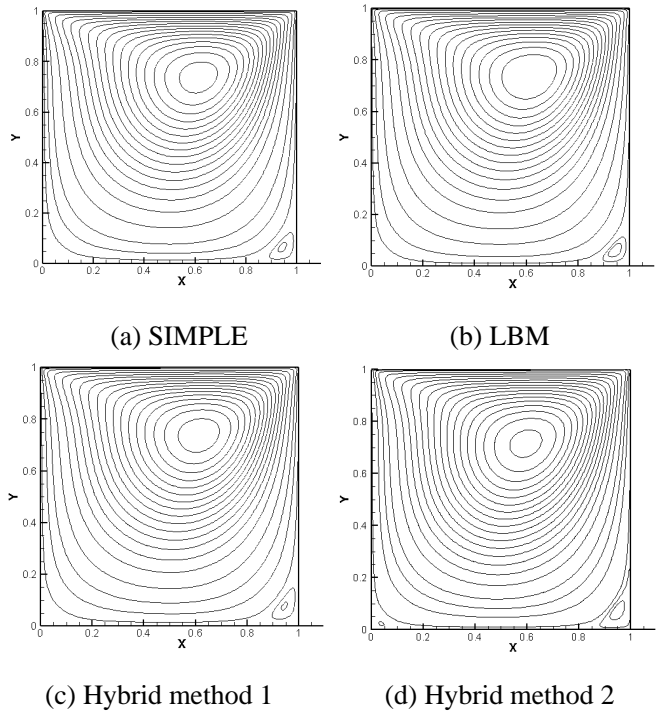

(a) SIMPLE                    (b) LBM

(c) Hybrid method 1        (d) Hybrid method 2

Figure 11 Streamlines at *Re* = 100 [36]

The lid driven flow is widely used as a benchmark solution to test the accuracy of a numerical method. To assess the hybrid methods, the results of Ref. [35] are used for comparison. The lid-driven flow is solved for three different Reynolds numbers at 100, 400, and 1000, respectively. Pure SIMPLE with QUICK scheme, LBM with nonequilibrium extrapolation boundary method, and the two methods combining SIMPLE and LBM are applied to solve this problem. As discussed above, all these four methods use the same grid of $160 \times 160$, while the grid in the message passing zone in the hybrid method is $3 \times 160$.

Figures 11 to 13 show the streamlines at three Reynolds numbers obtained from the three methods, while Fig. 14 and 15 are the horizontal velocity profiles in the middle of the x-direction and vertical velocity profiles in the middle of the y-direction comparing with that in Ref. [35]. It should be pointed out that the velocities in the reference need to multiply by 0.1 because the non-dimensional processes are different.

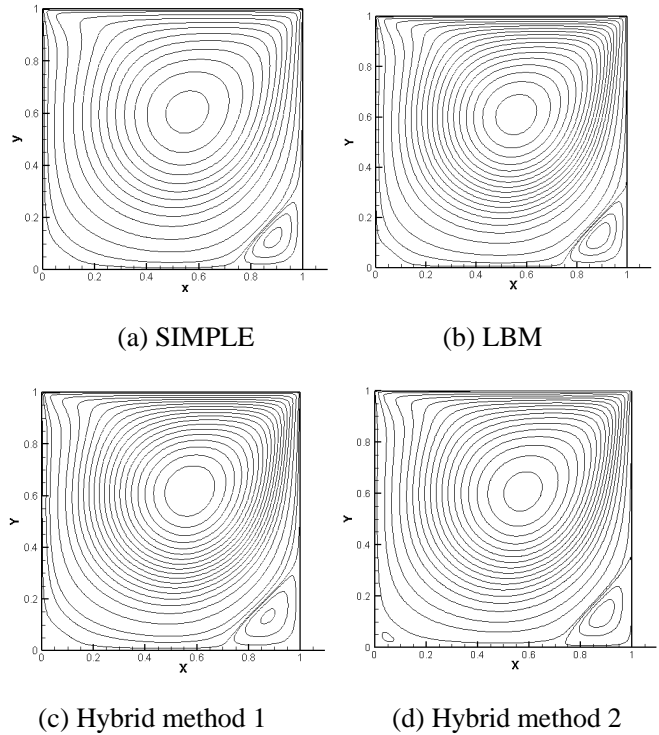

(a) SIMPLE                    (b) LBM

(c) Hybrid method 1           (d) Hybrid method 2

Figure 12 Streamlines at *Re* = 400 [36]

The streamlines obtained from the combined method are highly similar to that obtained from pure SIMPLE and pure LBM as shown in the Figs. 11 to 13. The positions of the centers of the primary vortices are (0.6125, 0.7375) by SIMPLE, (0.61875, 0.74375) by LBM, and (0.6172, 0.7344) in Ref. [35] at *Re*=100. And these three locations from these three sources are (0.5500, 0.60625), (0.55625, 0.6125), and (0.5547, 0.6055) for *Re*=400, and (0.5250, 0.55625), (0.53125, 0.56875), and (0.5313, 0.5625) for *Re*=1000. The differences of the locations the centers of the primary vortices are insignificant in three cases from SIMPLE and LBM. And the streamlines got from the two methods are highly similar

In addition to that, the horizontal velocity profiles in the middle of the *x*-direction and vertical velocity profiles in the middle of the y-direction obtained by SIMPLE and LBM are very close to that in reference as shown in Figs. 14 and 15. In other words, the SIMPLE and LBM used in this section are reliable. Due to this, the stream lines obtained from SIMPLE can be treated as standard results. Therefore, it can be concluded that the SIMPLE and LBM used in this section are reliable and the accuracy of the combine method only depends on the solution of the interface itself.

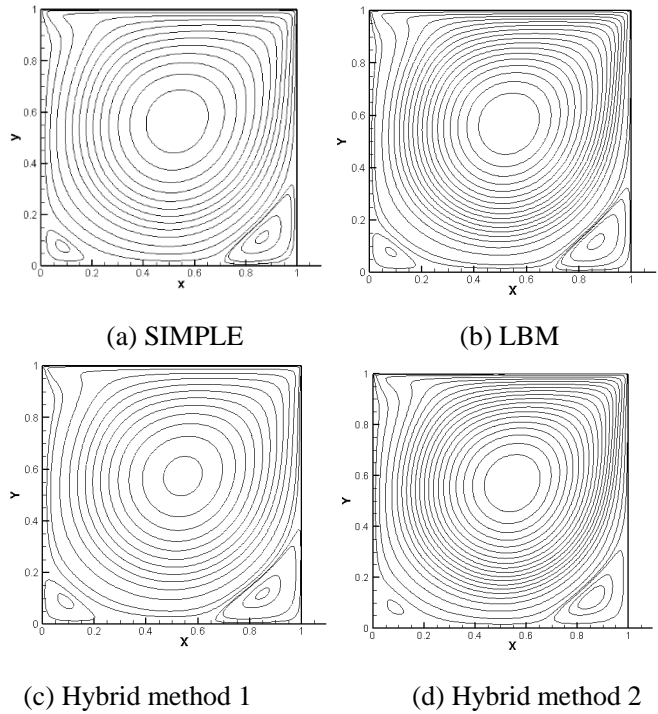

(a) SIMPLE                    (b) LBM

(c) Hybrid method 1          (d) Hybrid method 2

Figure 13 Streamlines at $Re = 1000$ [36]

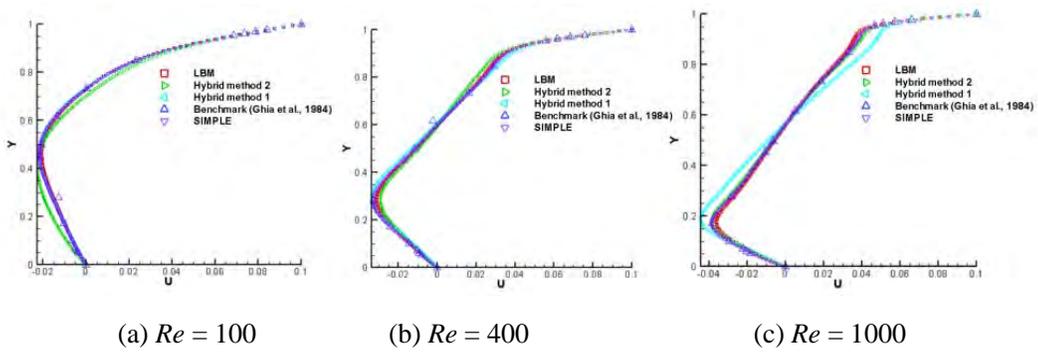

(a) $Re = 100$          (b) $Re = 400$          (c) $Re = 1000$

Figure 14 Horizontal velocity profiles

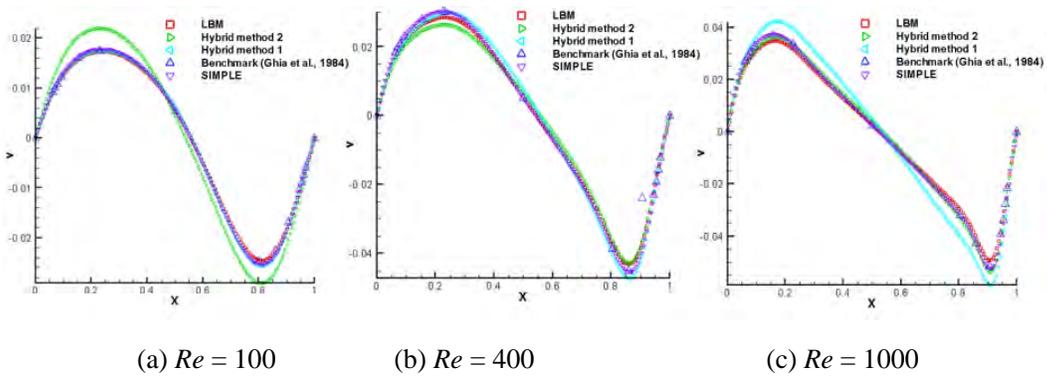

(a) $Re = 100$          (b) $Re = 400$          (c) $Re = 1000$

Figure 15 Vertical velocity profiles

On the other hand, the streamlines obtained from the hybrid method 1 and hybrid method 2 for Reynolds numbers at 100, 400 and 1000 shown in the Figs. 11 to 13 are almost the same as those obtained from SIMPLE and LBM. The positions of the centers of the primary vortices obtained from the hybrid method for three different Reynolds numbers are (0.6125, 07375), (0.5547, 0.6055) and (0.5313, 0.5625) for hybrid method 1 and (0.6000, 071875), (0.5750, 0.60625) and (0.54375, 0.56875) for hybrid method 2 respectively. Thus, the differences from the reference are still at the same level as that from pure SIMPLE and LBM. The hybrid method can reach the same flow pattern as the other two methods. Especially the results at the message passing zone did not show any instability.

When regarding the details of the fluid field, these two hybrid methods show different accuracies in different Reynolds numbers. The horizontal velocity profiles in the middle of the $x$-direction and vertical velocity profiles in the middle of the y-direction obtained by hybrid method 1 satisfy the result from Ref. [35] for Reynolds number equals 100 and 400. Meanwhile, the horizontal velocity profiles in the middle of the $x$-direction and vertical velocity profiles in the middle of the y-direction obtained by hybrid method 2 satisfy the result from Ref. [35] for Reynolds number equals 400 and 1000. Nonequilibrium extrapolation scheme in hybrid method 1 obtain the unknown density distribution by assuming the nonequilibrium part of the density distribution equals to that on the nearby inner nodes. So the bigger the velocity gradient is, the worse the accuracy is. By contrast, the finite-difference velocity gradient method in hybrid method 2 obtains the he unknown density distribution by relating the nonequilibrium part of the density distribution with the velocity distribution. So the bigger the velocity gradient is, the worse the accuracy is. Since the velocity gradient on the interface will become more valid with the increasing of Reynolds number. Hybrid method 1 is suitable for the case when Reynolds number is low while hybrid method 2 is suitable for the high Reynolds number case.

For the incompressible problem, LBM needs several times more computational time than that of the SIMPLE while LBM saves a great deal of computational time in the complex fluid flow problems [37]. The total computational time of any hybrid method always depends on the computational time of the slower one. The extra time consumption of message passing can be neglected comparing with the total computational time. Thus, the computational time in the lid-driven flow depends on that in the LBM zone when these two zones have the same grids. The computational efficiency of the hybrid methods are between those of SIMPLE and LBM. The main purpose to solve the lid-driven flow with the hybrid methods which reach the same accuracy with more time consuming comparing with SIMPLE is to certify these two hybrid methods can build a relation between LBM and FVM to solve the fluid flow problem together. These hybrid methods have the further to save time with the same accuracy for the problem including several parts that can take advantages of both SIMPLE and LBM in their subdomains.

# 4. A COUPLED LATTICE BOLTZMANN AND FINITE VOLUME METHOD

In Section 3, first strategy to fulfill LBM-FVM multiscale method for fluid flow simulation is discussed. Heat transfer plays an important role in many multiscale problems. A coupled LBM-FVM method for fluid flow and heat transfer is proposed in this section. After taking heat transfer in consideration, three more settings are needed in LBM-FVM hybrid method comparing with that in last chapter: temperature information transfer between subdomains; temperature interpolation due to the differences between FVM and LBM in computational nodes' locations; density information transfer between subdomains regarding temperature fields are highly related with density.

Nonequilibrium extrapolation scheme has been proved to be valid in combing LBM and FVM for the low speed fluid flow simulation in last chapter. Nature convection in this chapter satisfies this speed requirement. So nonequilibrium extrapolation scheme is employed to transfer the velocity and temperature information in this chapter. Natural convections in a squared enclosure with different Rayleigh numbers are solved using the coupled method and the results are compared with those obtained from pure LBM and pure FVM for validation of the combined method.

## 4.1 Combined LBM-FVM method for fluid flow and heat transfer problem

### 4.1.1 Problem statement

Natural convection of incompressible fluid in a squared enclosure as shown in Fig. 16 is used to test the coupled method. For the velocity field, non-slip condition is applied to all boundaries. The left boundary is kept at a constant temperature $T_h$ while the right boundary has a lower constant temperature of $T_l$. The top and bottom boundaries are adiabatic. Applying Boussinesq assumption, the problem can be described by the following governing equations:

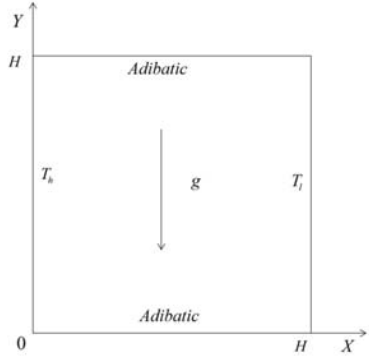

Figure 16 Physical model of the natural convection problem

$$\frac{\partial u}{\partial x} + \frac{\partial v}{\partial y} = 0 \tag{73}$$

$$\rho \left[ \frac{\partial u}{\partial t} + u \frac{\partial u}{\partial x} + v \frac{\partial u}{\partial y} \right] = -\frac{\partial p}{\partial x} + \mu \left( \frac{\partial^2 u}{\partial x^2} + \frac{\partial^2 u}{\partial y^2} \right) \tag{74}$$

$$\rho \left[ \frac{\partial v}{\partial t} + u \frac{\partial v}{\partial x} + v \frac{\partial v}{\partial y} \right] = -\frac{\partial p}{\partial y} + \mu \left( \frac{\partial^2 v}{\partial x^2} + \frac{\partial^2 v}{\partial y^2} \right) + \rho g \beta \left( T - T_l \right) \tag{75}$$

$$\left( \rho c_p \right) \left[ \frac{\partial T}{\partial t} + u \frac{\partial T}{\partial x} + v \frac{\partial T}{\partial y} \right] = k \left( \frac{\partial^2 T}{\partial x^2} + \frac{\partial^2 T}{\partial y^2} \right) \tag{76}$$

Equations (73) – (76) are subject to the following boundary and initial conditions:

$$x = 0, \ u = 0, \ v = 0, \ T = T_h \tag{77}$$

$$x = H, \ u = 0, \ v = 0, \ T = T_l \tag{78}$$

$$y = 0, \ u = 0, \ v = 0, \ \partial T / \partial y = 0 \tag{79}$$

$$y = H, \ u = 0, \ v = 0, \ \partial T / \partial y = 0 \tag{80}$$

Regarding Eq. (18) in the view of Chapman-Enskog expansion, Eq. (76) can be obtained from the D2Q9 model in LBM when effective gravity acceleration $G$ is defined as:

$$G = -\beta \left( T - T_l \right) g \tag{81}$$

where $\beta$ is the volume expansion coefficient of the fluid.

The SIMPLE algorithm with QUICK scheme [6] is employed to solve Eqs. (73) - (76). Prandtl number, $Pr$, and Rayleigh number, $Ra$, are the two non-dimensional parameters governing the natural convection.

$$Pr = \frac{\alpha}{\nu} \tag{82}$$

$$Ra = \frac{g\beta(T_h - T_l)H^3 Pr}{\nu^2} \tag{83}$$

For LBM Mach number, $Ma$, is needed:

$$Ma = \frac{u_c}{c_s} \tag{84}$$

where $u_c$ is the speed of sound that equals $\sqrt{g\beta(T_h - T_l)H}$. Since the natural convection in consideration is incompressible, $Ma$ can be any number in the incompressible region.

Applying the following non-dimensional variables

$$X = \frac{x}{H}, Y = \frac{y}{H}, U = \frac{u}{\sqrt{3}c_s}, V = \frac{v}{\sqrt{3}c_s}$$
$$\tau = \frac{t \cdot \sqrt{3}c_s}{H}, \theta = \frac{T - T_l}{T_h - T_l}, P = \frac{p}{3\rho c_s^2} \tag{85}$$

to Eqs. (73)-(80), the dimensionless governing equations are obtained:

$$\frac{\partial U}{\partial X} + \frac{\partial V}{\partial Y} = 0 \tag{86}$$

$$\frac{\partial U}{\partial \tau} + U\frac{\partial U}{\partial X} + V\frac{\partial U}{\partial Y} = -\frac{\partial P}{\partial X} + Ma\sqrt{\frac{Pr}{3Ra}}\left(\frac{\partial^2 U}{\partial X^2} + \frac{\partial^2 U}{\partial Y^2}\right) \tag{87}$$

$$\frac{\partial V}{\partial \tau} + U\frac{\partial V}{\partial X} + V\frac{\partial V}{\partial Y} = -\frac{\partial P}{\partial Y} + Ma\sqrt{\frac{Pr}{3Ra}}\left(\frac{\partial^2 V}{\partial X^2} + \frac{\partial^2 V}{\partial Y^2}\right) + \frac{Ma^2\theta}{3} \tag{88}$$

$$\frac{\partial \theta}{\partial \tau} + U\frac{\partial \theta}{\partial X} + V\frac{\partial \theta}{\partial Y} = Ma\sqrt{\frac{1}{3Ra \cdot Pr}}\left(\frac{\partial^2 \theta}{\partial X^2} + \frac{\partial^2 \theta}{\partial Y^2}\right) \tag{89}$$

For the heat transfer at the left boundary, Nusselt number, $Nu$, can be obtained by the nondimensional temperature gradient at the surface:

$$Nu = -\frac{\partial \theta}{\partial X}\bigg|_{X=0} \tag{90}$$

which reflects the ratio of convection to the conduction heat transfer across the wall.

### 4.1.2 Description of the Coupled Method

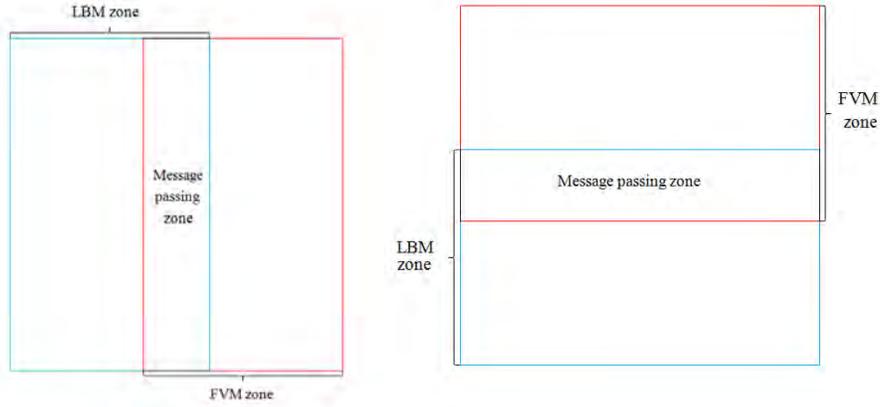

(a) Divided vertically              (b) Divided horizontally

Figure 17 Computational domains for LBM and FVM

This coupled method is designed to solve a single problem with FVM and LBM simultaneously. The computational domain is divided into LBM and FVM zones, and there is a public area between these two zones. The artificial boundary of FVM zone is the inner nodes of LBM zone while the LBM artificial boundary is inside the FVM zone. Two kinds of geometry settings are applied to test this coupled method shown in Fig. 17.

To fulfill the coupled method, the information on the artificial boundary needs to be obtained from the other subdomain. For the FVM zone, the velocity and temperature on the artificial boundary are needed from LBM zone. Density is not needed because FVM is solving the incompressible flow. The pressure on that boundary can be obtained directly from the FVM zone itself [6]. On the other hand, LBM needs the velocity, temperature and the density information of the artificial boundary from the FVM zone. It is not straightforward to transfer the density from an incompressible FVM zone to the compressible LBM zone. The average density in the message passing zone, $\rho_0$, is calculated in the message LBM zone, and the FVM zone provides the average pressure in the message passing zone, $\bar{p}$. Meanwhile, the pressure on the LBM artificial boundary, $p^L$, can be obtained from the FVM zone pressure, $p^S$. It is shown that there is very small difference between $p^L$ and $p^S$ in the message passing zone. But this small difference leads evident error when I tried to fulfill this combine method. It is also found that the pressure gradients differences from the two methods in the message passing zone is not evident either, Then the difference between $\bar{p}^L$ and $\bar{p}$ is similar as that between $p^L$ and $p^S$ where $\bar{p}^L$ is average pressure in the message passing zone calculated by the LBM zone results. Therefore it is reasonable to assume that

$$p^L - \bar{p}^L = p^S - \bar{p} \tag{91}$$

Regarding the relation between pressure and density in LBM, it can be obtained that

$$\rho^L c_s^2 - \rho_0 c_s^2 = p^S - \bar{p} \tag{92}$$

Then the unknown $\rho^L$ can be obtained that

$$\rho^L = \rho_0 \left( \frac{p^S - \bar{p}}{\rho_0 c_s^2} + 1 \right) \tag{93}$$

The density information on the artificial boundary of LBM can be obtained by the result of the FVM result. Staggered grid is applied to SIMPLE algorithm.

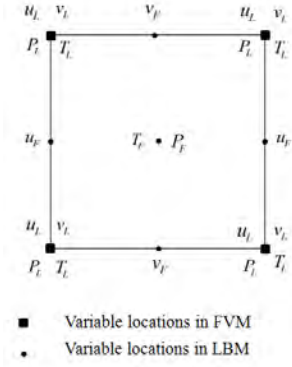

Figure 18 Variables locations

Figure 18 shows the locations of variables in a control volume for LBM and FVM. The variables in LBM all locates on the corner of the control volume while velocity, pressure and temperature in FVM have different locations in the control volume. Central difference is applied to message passing processes due to the variable location differences.

After transferring the information from each other, the FVM and LBM zones need to be solved independently for each time step. For the FVM zone, temperatures and velocities on the four boundaries are known so that the solution procedure is straightforward. There are several choices for the LBM boundary conditions. Nonequilibrium extrapolation scheme is applied to both velocity and temperature fields in the solving process. Assuming $x_b$ is the boundary node and $x_f$ is its nearby inner mode, the density and energy distribution at the artificial boundary are:

$$f_i(x_b, t) = f_i^{eq}(x_b, t) + f_i(x_f, t) - f_i^{eq}(x_f, t) \tag{94}$$

$$g_i(x_b, t) = g_i^{eq}(x_b, t) + g_i(x_f, t) - g_i^{eq}(x_f, t) \tag{95}$$

The temperatures on the boundaries are known in every time step, Eq. (95) can be applied to the thermal boundary conditions for Eq. (93). For the velocity field, the boundary density is only known on the artificial boundary. It is common to approximate the density on the fixed boundary by

$$\rho(x_b, t) = \rho(x_f, t) \tag{96}$$

## 4.2 Two geometry results comparison

Natural convection in a squared enclosure is solved for three different Rayleigh numbers at $10^4$, $10^5$ and $10^6$ while the Prantl number is kept at 0.71. Pure LBM and pure FVM are reported in the literature to be suitable for the natural convection in a cavity. Thus, these two methods are applied to solve the test cases. If the LBM results agree with the FVM results, it can be concluded that these results can be used as standard results for comparison. It can also verify that the codes for the two subdomains are reliable in the coupled method. Then only the message passing method between the two subdomains affects the results from the coupled methods. Two coupled methods with different geometry settings are applied. When the domain is divided vertically, it is referred to as Coupled Method 1. And the Coupled Method 2 divides the domain horizontally as shown in Fig. 17. Temperature field, streamline and Nusselt number on the left wall obtained from these four methods are compared for the three cases. Nondimensional variables defined in Eq. (85) are applied in the comparisons.

Figures 19 and 20 show comparisons of temperature field and streamlines obtained by different methods for the case that Rayleigh number is $10^4$. It is obvious that natural convection has dominated the heat transfer process and there is a stream line vertex near the center of the cavity. Temperature fields and streamlines obtained from pure LBM and pure FVM agree with each other well as shown in Fig. 19 and 20. In addition, there is not any noticeable difference between the results obtained from coupled methods 1 and 2 and the results of coupled methods agreed with that from the pure FVM and pure LBM very well.

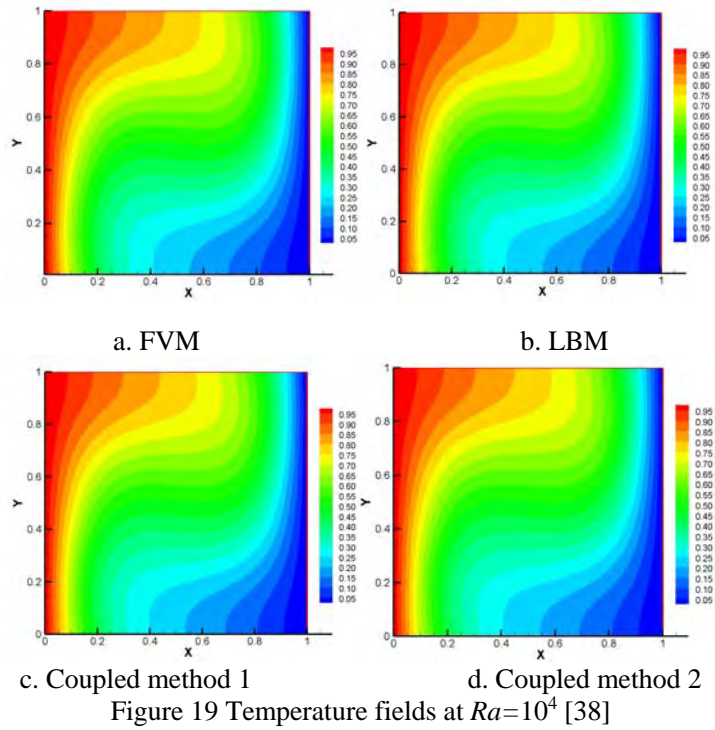

a. FVM                                b. LBM

c. Coupled method 1                   d. Coupled method 2

Figure 19 Temperature fields at $Ra=10^4$ [38]

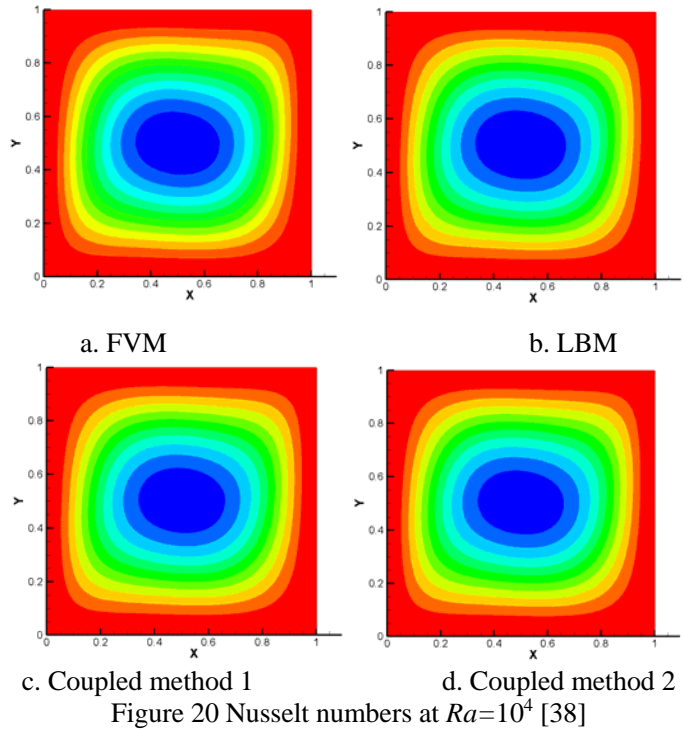

a. FVM                                b. LBM

c. Coupled method 1                   d. Coupled method 2

Figure 20 Nusselt numbers at $Ra=10^4$ [38]

Figure 21 shows the Nusselt number at heated wall along the vertical direction of the enclosure obtained from different methods. There is a little difference between the Nusselt numbers obtained from pure LBM and pure FVM. The cause of this difference is that FVM is based on the incompressible fluid assumption while LBM is based on compressible fluid assumption. Meanwhile the Nusselt number tendencies of the two coupled methods are very close to that of the two pure methods.

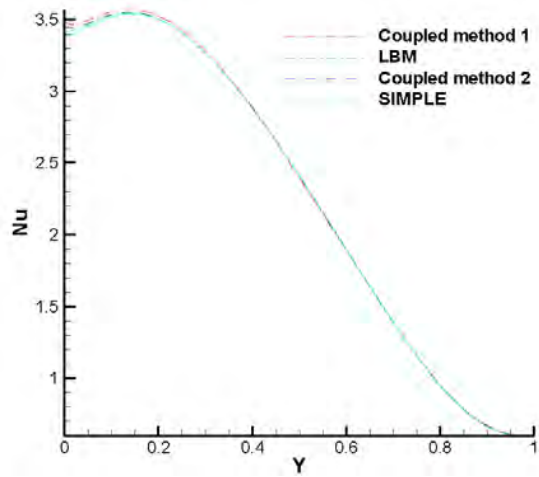

Figure 21 Nusselt numbers at $Ra=10^4$ [38]

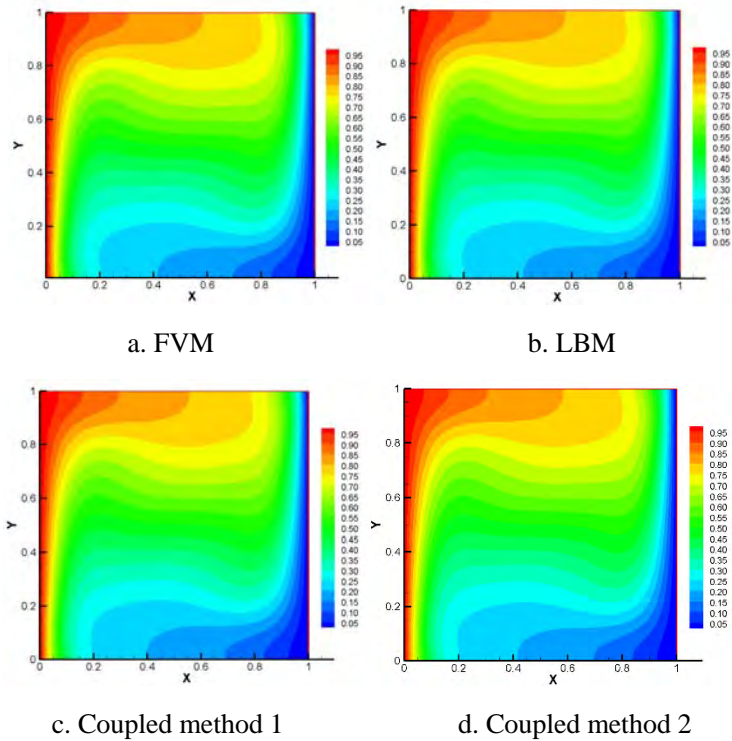

a. FVM                                    b. LBM

c. Coupled method 1              d. Coupled method 2

Figure 22 Temperature fields at $Ra=10^5$ [38]

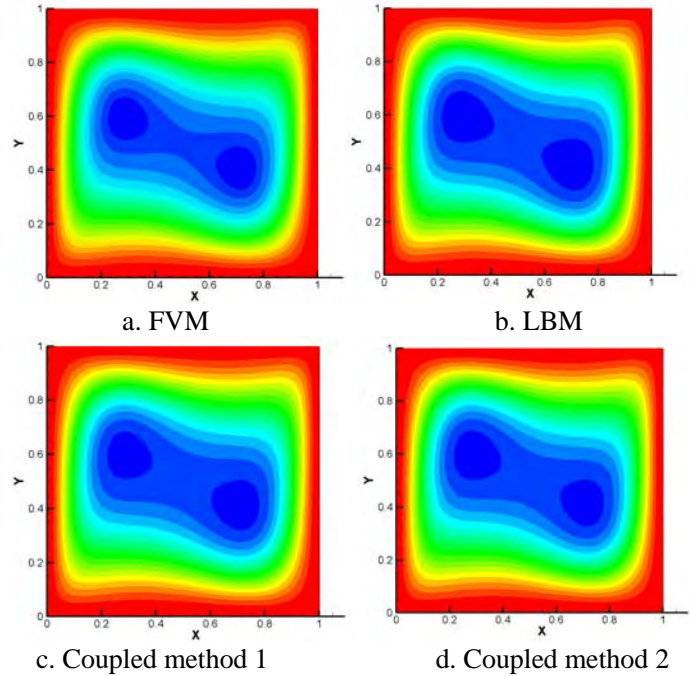

a. FVM                                    b. LBM

c. Coupled method 1                       d. Coupled method 2

Figure 23 Streamlines at $Ra$=$10^5$ [38]

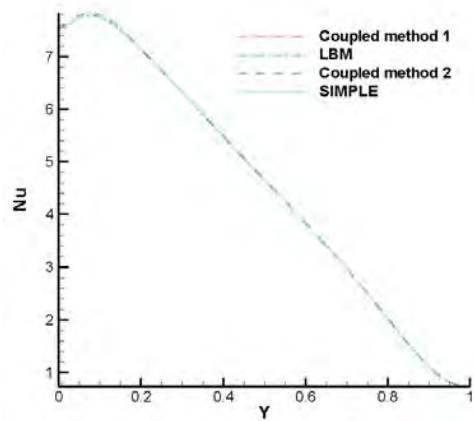

Figure 24 Nusselt numbers at $Ra$=$10^5$ [38]

Figures 22 and 23 show ther temperature field and streamlines obtained by different methods for the case that Rayleigh number is $10^5$. It can be see that the pure FVM and pure LBM reach similar temperature fields and streamlines. The convection effect becomes more pronounced as the Rayleigh number increases. Two vertexes appear and the temperature gradients near the vertical boundary increase. The two coupled method results still agree very well with that in the pure methods as shown in Figs. 22 and 23. Meanwhile, Fig. 24 shows that the difference between Nusselt numbers obtained from pure FVM and pure LBM is larger than that in Fig. 21; but the largest difference is still round 2%. The Nusselt numbers from the two coupled methods are closer to the results of pure LBM than that of the pure FVM. Since both coupled methods 1 and 2 have half regions with LBM that do not have incompressible fluid assumption, the fluid in the entire computational domain of the coupled methods can be considered as compressible. The Nusselt number differences between the two coupled methods are not larger than that between the two pure methods.

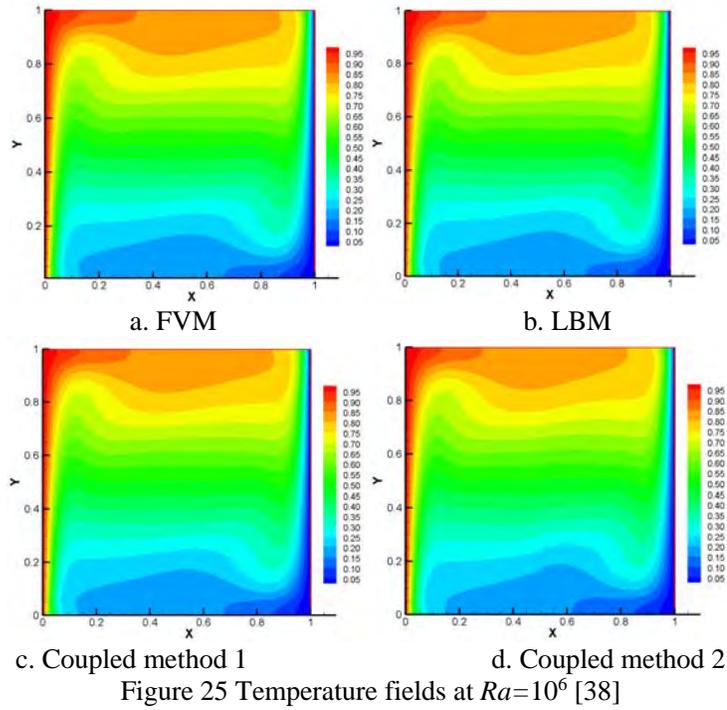

a. FVM                                    b. LBM

c. Coupled method 1                       d. Coupled method 2

Figure 25 Temperature fields at $Ra=10^6$ [38]

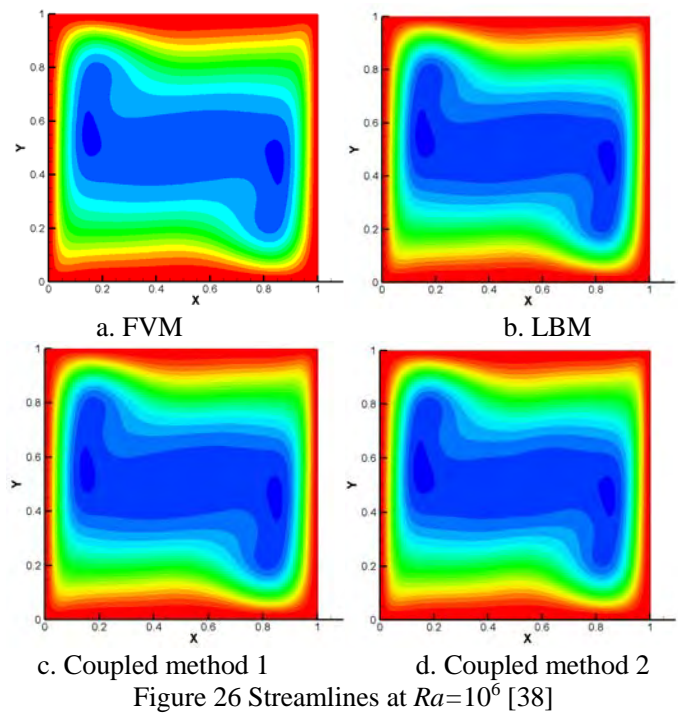

a. FVM                                    b. LBM

c. Coupled method 1                       d. Coupled method 2

Figure 26 Streamlines at $Ra=10^6$ [38]

Convection continues to become stronger when Rayleigh number is increased to $10^6$. Figures 25 and 26 show that all four methods yield the similar temperature fields and streamlines. There are still two independent stream line vertexes that are closer to the vertical boundaries; this indicates a stronger convection effect comparing with the results when Rayleigh number is $10^5$. As for the Nusselt number, Fig. 27 shows that the results from the coupled methods 1 and 2 are very close to that from the pure LBM. And the difference between the two coupled methods and pure FVM is still acceptable.

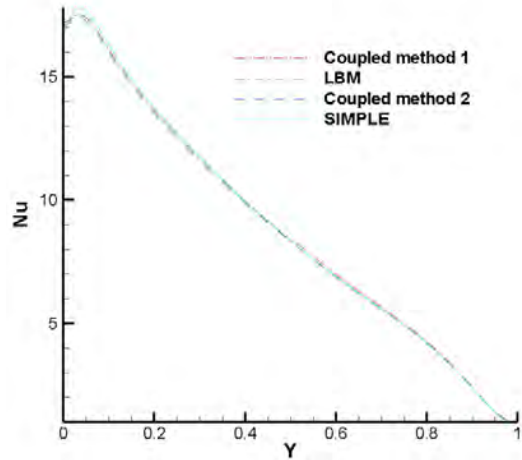

Figure 27 Nusselt numbers at $Ra=10^6$ [38]

Therefore, the results obtained from the four methods agreed with each other very well at different Rayleigh numbers of $10^4$, $10^5$ and $10^6$. The geometric settings do not affect the accuracy of the coupled method. The results of this work demonstrated that the coupled method is reliable to solve natural convection problems.

## 5. A HYBRID LATTICE BOLTZMANN AND FINITE VOLUME METHOD

In Sections 3 and 4, LBM-FVM multiscale method was proposed to solve fluid flow and heat transfer problems. The whole domain is divided into two subdomains and these subdomain are solved with LBM and FVM respectively. This strategy of hybrid method is suitable for simulating problems involved with multiple length scales.

In another strategy of hybrid LBM-FVM method, velocity, pressure and density are obtained from the density distribution solved by LBM, while the temperature field can be obtained directly using FVM based on the other solved macroscopic variables from LBM. This is a promising numerical method since the main advantages of LBM lies in obtaining velocity field. Two addition settings are needed to fulfill this LBM-FVM hybrid method: the variable locations in LBM and FVM are different and proper difference is needed in the simulation process; FVM and LBM are different time scale methods and time steps are set different in LBM and FVM.

Pure thermal LBM, pure FVM, and hybrid LBM-FVM are employed to solve the natural convection in a fixed-boundary cavity with different Rayleigh numbers. The results are compared with reference ones for validation.

### 5.1 Coupled LBM-FVM approach

The velocity field is obtained by the LBM while FVM solves the temperature field in the hybrid method. They have been included in Section 2. Different from the multiscale method in Chapters 3 and 4, the LBM-FVM hybrid method in this chapter apply LBM and FVM to the whole computing domain. The locations for the temperature and velocity in the control volume are shown in Fig. 28.

A corrected method to solve convective-diffusion equation based on SIMPLE was proposed in reference [39]. Based on this result, a LBM-FVM hybrid method is proposed in this chapter. The temperature field can be solved by energy balance in the control volume after the velocity field is obtained from the LBM.

Assume the grid in LBM is $2NX \times 2NY$ after mesh independent testing for the LBM. In the natural convection cases solved in this section, $160 \times 160$ is the selected mesh. It is quite straightforward to assume that the grid for the temperature field is the same as that in the velocity field shown in Fig. 29. The velocities on the faces of control volume that are needed in the FVM are obtained by the interpolating of the LBM results. Figure 30 shows another method to set the grid for FVM that $NX \times NY$ grid is applied to FVM for the temperature field.

The dashed lines show the new control volume for the temperature field. The obtained velocity results can be applied to FVM directly without interpolation. Then the temperature on the computing nodes of can be obtained by the interpolation of the FVM results for these two grids setting method, and they can reach the same results after comparing the results. Since the computational time for the temperature field in the second method is much shorter than that of the first one, the second method is preferred in the hybrid method.

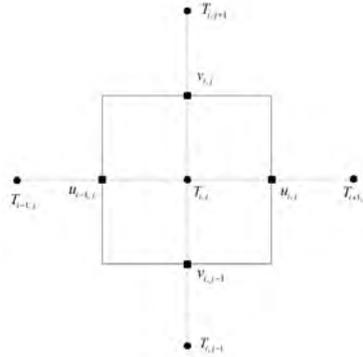

Figure 28 Control volume in FVM

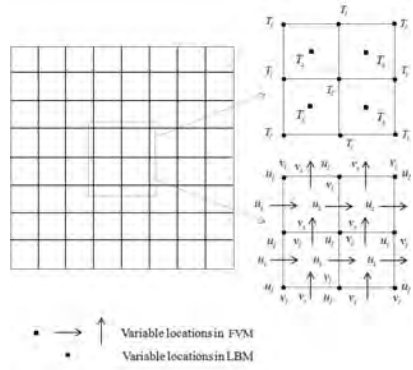

Figure 29 Grid setting 1

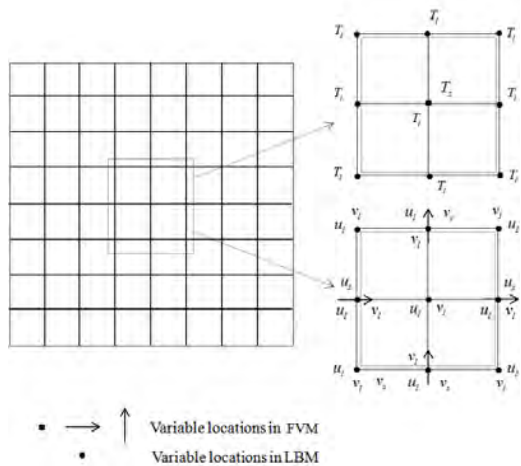

Figure 30 Grid setting 2

Non-dimensional process in FVM is the same as that in Section 4 and $2NX \times 2NY$ is employed in LBM. Correspondently, time steps in FVM and LBM are $H/\left(\sqrt{3}c_s\right)$ and $H/\left(2NX \cdot \sqrt{3}c_s\right)$ respectively. Therefore, in every FVM time step, LBM runs $2NX$ time steps.

## 5.2 Results and discussions

Natural convection in squared enclosure is widely used as a benchmark problem for validation of the numerical methods. Pure LBM, pure FVM and the hybrid method are applied to simulate the natural convection independently for three different Rayleigh numbers at $10^4$, $10^5$ and $10^6$, while the Prantl number is 0.71. It is reported by several groups independently that pure LBM and pure FVM are both suitable for the natural convection in this range [40, 41]. So when the results of these two methods agree with each other well, they can be treated as the benchmark results for comparison.

Under the Boussinesq assumption, variation of density with temperature is considered in the buoyancy force. In the gravitational field, there is a tendency that lighter fluid goes upward and the heavier one goes downward. The boundary layer of a vertical wall with high constant temperature turns thicker with the increasing of the height, while the boundary layer of a vertical wall with low constant temperature becomes thicker with the decreasing of the height. So the fluid near the left wall moves up but the fluid near the right wall moves down. The closer to the left boundary, the hotter the fluid is since the left boundary has the highest temperature. Therefore the upward moving fluid near the left moves towards right near the top of the cavity and the downward moving fluid moves towards left near the bottom of the cavity. And the two flows combine together as a vortex. For the case that Rayleigh number is equal to $10^4$, the streamlines and temperature fields obtained by the pure FVM and pure LBM agreed with each other well as shown in Figs. 31 and 32, respectively. It is shown that the natural convection has governed the heat transfer process. There is a streamline vertex in the center of the cavity and the temperature field also shows the character of the convection. The streamlines and temperature field obtained from the hybrid method matches the results from the other two methods very well.

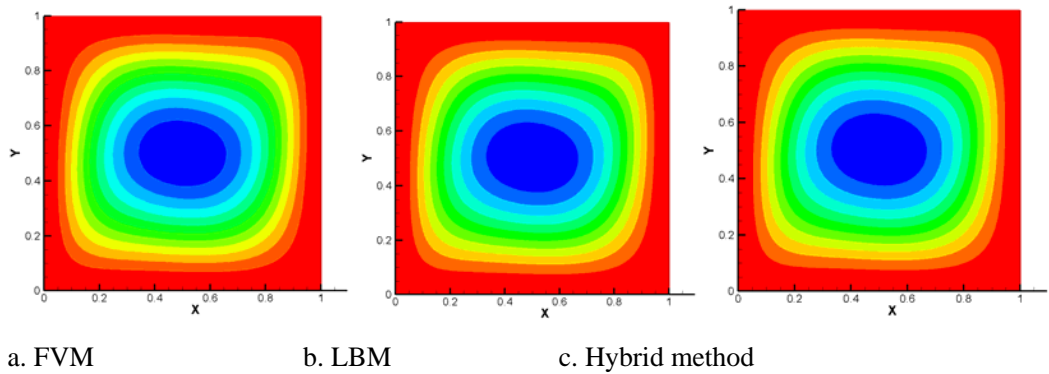

a. FVM                         b. LBM                    c. Hybrid method

Figure 31 Streamlines comparison at $Ra$=$10^4$ [42]

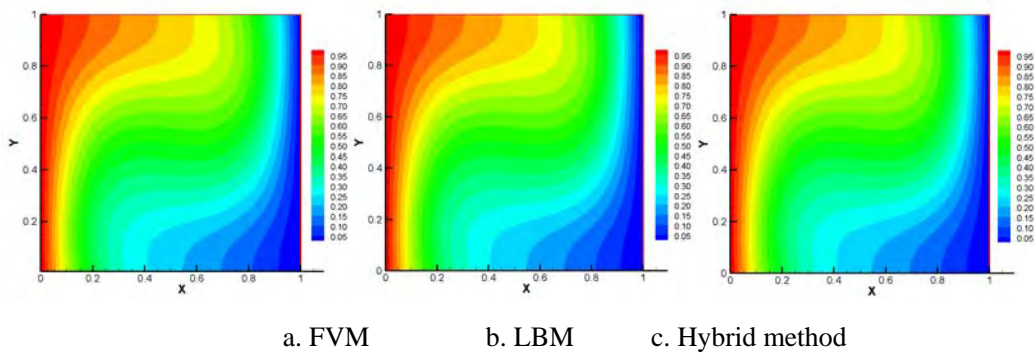

a. FVM               b. LBM          c. Hybrid method

Figure 32 Temperature field comparison at $Ra$=$10^4$ [42]

The natural convection effect becomes more obvious when the Rayleigh number is increased to $10^5$. Figures 33 and 34 show that the streamlines and temperature fields from the three methods agree with each other very well. There are two vertexes in the flow field and the temperature gradient at the left and right boundaries is larger than that for the case of $Ra = 10^4$. The differences of the streamlines and temperature fields among the three methods are still not noticeable when the Rayleigh number grows to $10^6$ as shown in Figs. 35 and 36. The two streamline vertexes locate farther from the cavity center and the temperature changes turn closer to the left and right boundaries. So the streamlines and temperature fields of the hybrid method agree with those obtained by the other two methods for the three Rayleigh numbers.

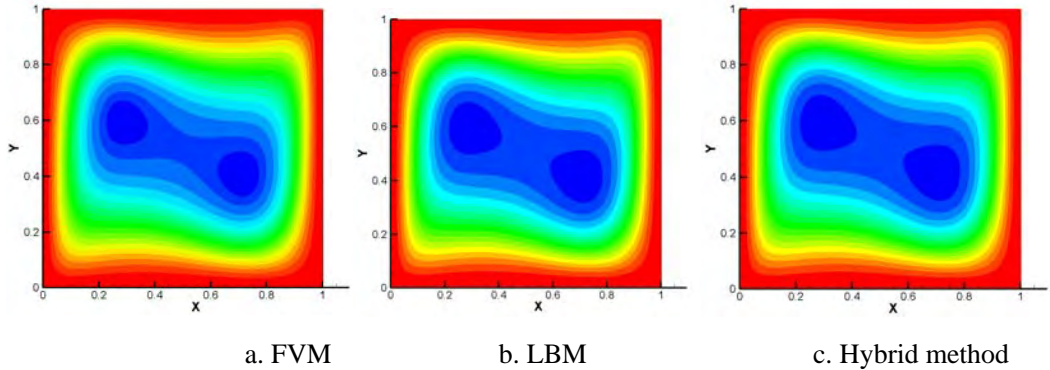

a. FVM        b. LBM        c. Hybrid method

Figure 33 Streamlines comparison at $Ra=10^5$ [42]

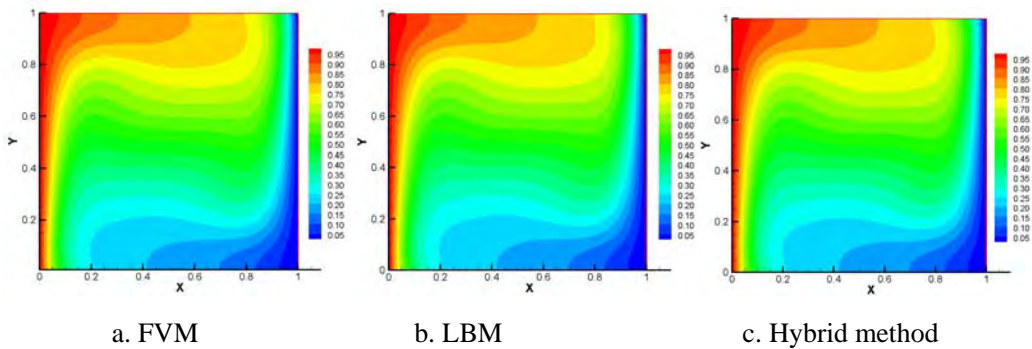

a. FVM        b. LBM        c. Hybrid method

Figure 34 Temperature field comparison at $Ra=10^5$ [42]

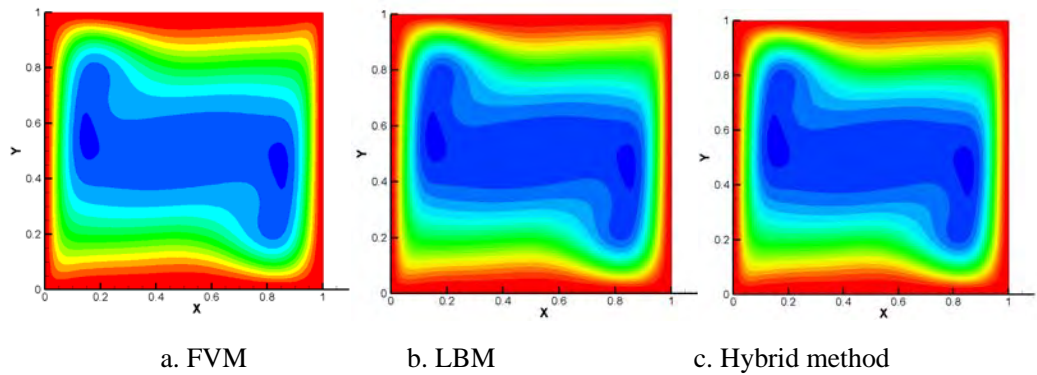

a. FVM        b. LBM        c. Hybrid method

Figure 35 Streamlines comparison at $Ra=10^6$ [42]

Comparison of Nusselt number obtained from the three methods is shown in the Fig. 5-10 for different cases. It can be seen that the results agree with each other very well in the three cases. The main difference locates at the maximum Nusselt number part, which is more obvious as the Rayleigh number increases. Meanwhile Table 1 and 2 give the comparisons of the maximum Nusselt number and its location for the three methods and the results from Ref. [41]. For the maximum Nusselt number, the hybrid result is closest to the reference result in the first two cases while the pure LBM has the closest result when Rayleigh number equals $10^6$. The hybrid method result is closest to the reference result in all the three cases. The largest difference between all the three methods results and the reference result is still under 3% which is acceptable.

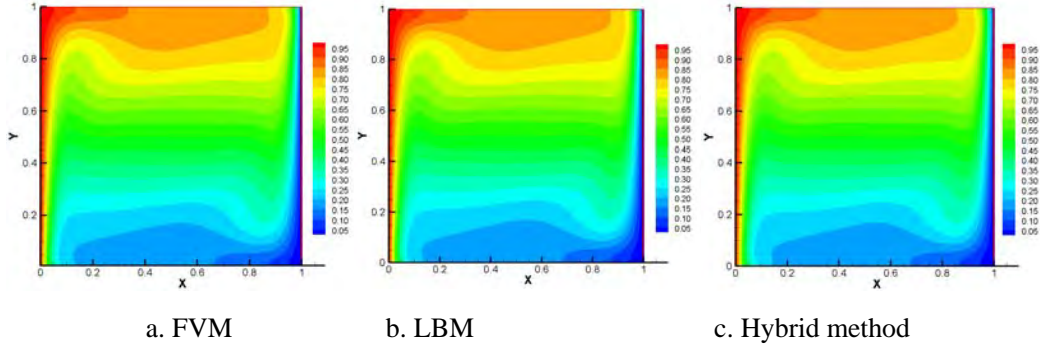

a. FVM                        b. LBM                        c. Hybrid method

Figure 36 Temperature field comparison at $Ra$=$10^6$ [42]

Table 1 Comparison of the maximum Nusselt numbers [42]

|               | $Ra$=$10^4$ | $Ra$=$10^5$ | $Ra$=$10^6$ |
|---------------|-------------|-------------|-------------|
| Hybrid method | 3.5324      | 7.6970      | 17.3354     |
| FVM           | 3.5486      | 7.8382      | 17.8399     |
| LBM           | 3.5481      | 7.7907      | 17.5133     |
| Reference [41]| 3.5309      | 7.7201      | 17.5360     |

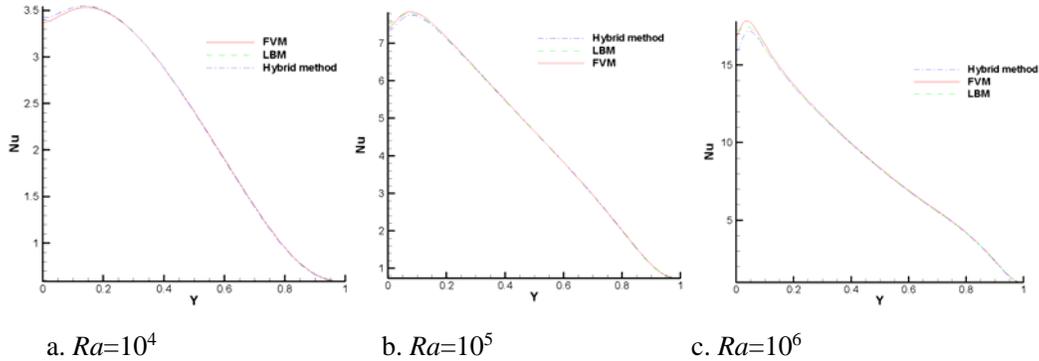

a. $Ra$=$10^4$                 b. $Ra$=$10^5$                 c. $Ra$=$10^6$

Figure 37 Comparison of Nusselt numbers [42]

Table 2 Comparison of the locations of maximum Nusselt numbers

|               | $Ra$=$10^4$ | $Ra$=$10^5$ | $Ra$=$10^6$ |
|---------------|-------------|-------------|-------------|
| Hybrid method | 0.1437      | 0.0812      | 0.0437      |
| FVM           | 0.1375      | 0.0813      | 0.0334      |
| LBM           | 0.1375      | 0.0750      | 0.0313      |
| Reference [41]| 0.1439      | 0.0820      | 0.0392      |

The streamlines and temperature fields obtained by the three methods agree with each other very well for different Rayleigh numbers at $10^4$, $10^5$ and $10^6$. The hybrid method also has a good accuracy for the Nusselt number when comparing with the reference and the two pure methods results. Thus, the hybrid LBM-FVM is reliable for the natural convection simulation.

# 6. A COMBINED LATTICE BOLTZMANN AND MONTE CARLO METHOD

To take advantages of both LBM and FVM, a LBM-FVM hybrid method was proposed and verified in Section 5. Monte Carlo method has valid advantages in solving some heat transfer problems. LBM-MCM combined method is proposed for fluid flow and heat transfer simulation in this section. Velocity field is solved with LBM while temperature is obtained using MCM. Natural convection with different Rayleigh numbers are solved to verify this numerical method.

## 6.1 Monte Carlo Method (MCM) for heat transfer

A rectangular grid system of mesh size $\Delta x \times \Delta y$ is selected for the two-dimensional computing domain shown in Fig. 38. The inner computing node $(i,j)$ relates to its surrounding nodes by:

$$T_{i,j} = P_{x+}T_{i+1,j} + P_{x-}T_{i-1,j} + P_{y+}T_{i,j+1} + P_{y-}T_{i,j-1} \tag{97}$$

where the possibilities $P_{x+}$, $P_{x-}$, $P_{y+}$ and $P_{y-}$ are positive; their sum has to be 1.

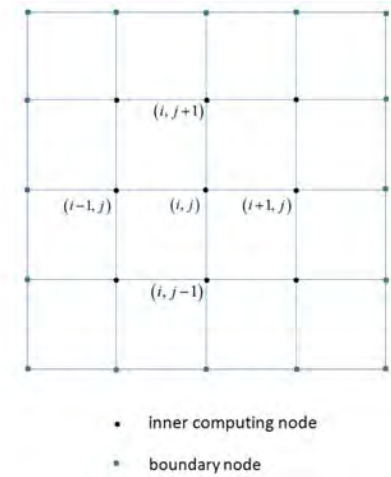

Figure 38 Computing grid

For the conduction heat transfer problem, these possibilities are:

$$\begin{cases} P_{x+} = P_{x-} = \dfrac{\Delta y / \Delta x}{2(\Delta y / \Delta x + \Delta x / \Delta y)} \\ P_{y+} = P_{y-} = \dfrac{\Delta x / \Delta y}{2(\Delta y / \Delta x + \Delta x / \Delta y)} \end{cases} \tag{98}$$

$$\begin{cases} P_{x+} = \dfrac{\alpha \Delta y / \Delta x - u \Delta y}{D} \quad P_{x-} = \dfrac{\alpha \Delta y / \Delta x}{D} \\ P_{y+} = \dfrac{\alpha \Delta x / \Delta y - v \Delta x}{D} \quad P_{y+} = \dfrac{\alpha \Delta x / \Delta y}{D} \end{cases} \tag{99}$$

where

$$D = 2(\Delta y / \Delta x + \Delta x / \Delta y) - u \Delta y - v \Delta x \tag{100}$$

For the convection heat transfer problem, the horizontal velocity $u$ and vertical velocity $v$ have effects on $P_{x+}$, $P_{x-}$, $P_{y+}$ and $P_{y-}$. The possibilities should be modified to the following equation:

The statistical procedure MCM uses random walkers to solve the heat transfer problem [43]. A random walker locates at node $(i,j)$ at beginning and a random number $RN$ is chosen in the uniformly distribution set from 0 to 1. This random walker will change its position by the following rules:

$$\begin{cases} \textit{if } 0 < RN < P_{x+} & \textit{from}(i,j)\textit{ to}(i+1,j) \\ \textit{if } P_{x+} < RN < P_{x+} + P_{y+} & \textit{from}(i,j)\textit{ to}(i,j+1) \\ \textit{if } P_{x+} + P_{y+} < RN < P_{x+} + P_{y+} + P_{x-} & \textit{from}(i,j)\textit{ to}(i-1,j) \\ \textit{if } 1 - P_{y-1} < RN < 1 & \textit{from}(i,j)\textit{ to}(i,j-1) \end{cases} \qquad (101)$$

Once the random walker has completed its first step, the procedure continues for the second step. This process goes on till that random walker reaches the boundary. Then the boundary condition $T_w(1)$ is recorded for the inner node $(i,j)$. This process is repeated $N$-1 times and the recorded boundary conditions are $T_w(2)$ to $T_w(N)$. With these $N$ results, the MCM estimation for $T(i,j)$ can be expressed as:

$$T(i,j) = \frac{1}{N}\sum_{n=1}^{N} T_w(n) \qquad (102)$$

Then all the inner computing nodes can be obtained by this method. The treatments to different kind boundaries can be found in reference [43].

### 6.2 Coupled LBM-MCM method

A combined LBM-MCM method is designed to solve a fluid flow and heat transfer problem with LBM and MCM simultaneously. Same grid system is used in both LBM and MCM. The velocity field is obtained by LBM and MCM solves the temperature field.

Figure 39 shows the flowchart for the combined method. For the natural convection problem in consideration, temperature and velocity have effects on each other. Therefore this method needs to run LBM and MCM in consequence for each step. After setting the initial condition, LBM solves the velocity field with the initial temperature field. Then the temperature field is obtained by MCM with the LBM velocity result. This temperature can be applied to LBM in next step. This process is repeated till converged results are reached.

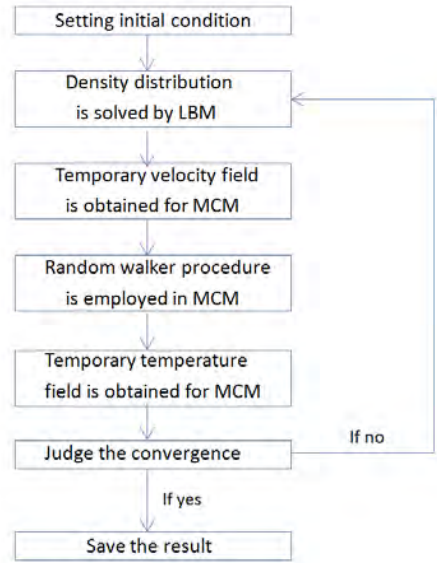

Figure 39 Flowchart for combined LBM-MCM

### 6.3 Numerical methods validation

### 6.3.1 Validation of MCM

In Section 3, LBM has been proved to be valid in solving incompressible fluid flow problems. In this section, MCM is verified for heat transfer simulation. As discussed in section 6.1, MCM has the same approach in solving conduction and convection heat transfer problems. A pure conduction problem is solved to test the MCM. In Fig. 40, the left, right and bottom of the two-dimensional domain are kept at $T_c$ and the top boundary has a higher temperature of $T_h$. It is assumed that the thermal conductivity is independent from the temperature. The energy equation for this problem is

$$\frac{\partial^2 T}{\partial x^2} + \frac{\partial^2 T}{\partial y^2} = 0, \quad 0 < x < H, 0 < y < H \tag{103}$$

with the following boundary conditions:

$$T = T_c, \quad x = 0 \; or \; H, 0 < y < H \tag{104}$$

$$T = T_c, \quad y = 0, 0 < x < H \tag{105}$$

$$T = T_h, \quad y = H, 0 < x < H \tag{106}$$

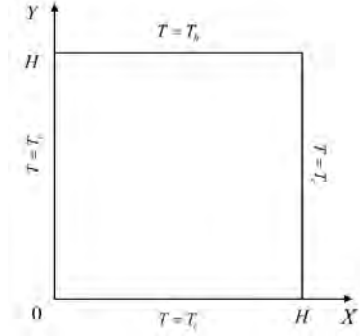

Figure 40 Two-dimensional steady state heat conduction

Defining the following non-dimensional variables

$$\begin{cases} X = \dfrac{x}{H}, Y = \dfrac{y}{H}, \\ \theta = \dfrac{T - T_c}{T_h - T_c} \end{cases} \tag{107}$$

Equations. (103)-(106) become

$$\frac{\partial^2 \theta}{\partial X^2} + \frac{\partial^2 \theta}{\partial Y^2} = 0, \quad 0 < X < 1, 0 < Y < 1 \tag{108}$$

with the following boundary conditions:

$$\theta = 0, \quad X = 0 \; or \; 1, 0 < Y < 1 \tag{109}$$

$$\theta = 0, \quad Y = 0, 0 < X < 1 \tag{110}$$

$$\theta = 1, \quad Y = 1, 0 < X < 1 \tag{111}$$

This problem can be solved analytically by separation of variable method [44].

$$\theta = \frac{2}{\pi} \sum_{n=1}^{\infty} \frac{(-1)^{n+1}+1}{n} \frac{\sinh(n\pi Y)}{\sinh(n\pi)} \sin(n\pi X) \tag{112}$$

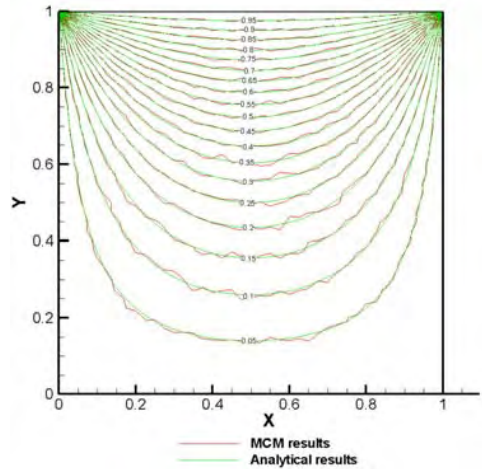

Figure 41 Conduction temperature field comparison [45]

Figure 41 shows the comparison of the analytical and MCM temperature fields. As discussed in section 6.2, MCM is a statistical method to simulate the heat transfer process. Its results are based on a large number of random walkers' results. The MCM isothermal lines are not that smooth due to its procedure nature. These two methods isothermal lines agree with each other well. Therefore, MCM in this section is valid for the heat transfer problem.

### 6.3.2 Natural convection in rectangular enclosure

In case 1, Figure 42 shows the streamlines for Case 1 that $Ra$ and $Pr$ are $10^4$ and $0.71$, respectively. There is a vortex in the cavity due to the convection effect. Figure 43 is the temperature field obtained from the combined method.

Both streamlines and temperature field agree with the benchmark solutions very well except the unsmoothness of some isothermal lines. As discussed in the pure conduction problem, the unsmoothness is led by the nature of MCM and its effect to the streamlines is insignificant. Meanwhile, Table 1 summarize maximum Nusselt number $Nu_{max}$, maximum Nusselt number location $Y_{Nu_{max}}$ and average Nusselt number $Nu_{ave}$ obtained from the LBM-MCM and Reference (Davis, 1983). It can be seen that the combined LBM-MCM method has a good agreement with benchmark standard solutions for these three parameters for Case 1.

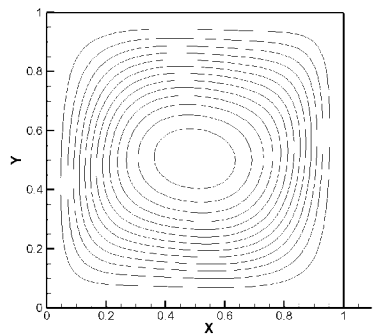

Figure 42 Streamlines in case 1 [45]

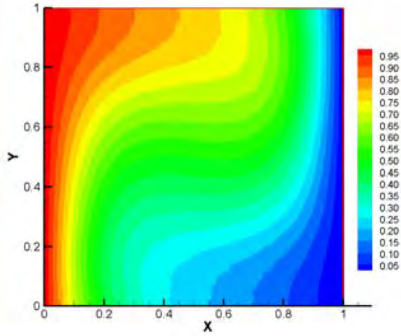

Figure 43 Temperature field in Case 1 [45]

Case 2 is also studied when $Ra$ grows to $10^5$ and $Pr$ is kept $0.71$. Convection plays a more important rule due to the increased $Ra$. Two vortexes locate in the cavity are shown in Fig. 44. And the temperature field in Fig. 45 also indicates a pronounced convection effect.

Table 3 also showed the comparison of $Nu_{max}$, $Y_{Nu_{max}}$ and $Nu_{ave}$ obtained from the present LBM-MCM with the benchmark solutions for this case. The streamlines, temperature filed and Nusselt number results in combined LBM-MCM method all agree with the benchmark solutions well. Similar to Case 1, the unsmooth isothermal lines effect is insignificant. Therefore the combined LBM-MCM method can give good predictions to these two natural convection cases.

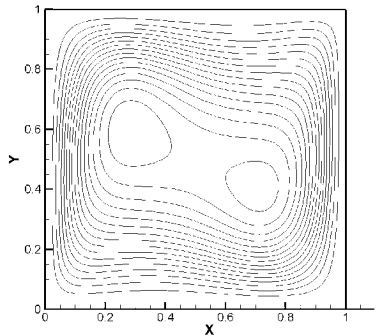

Figure 44 Streamlines in case 2 [45]

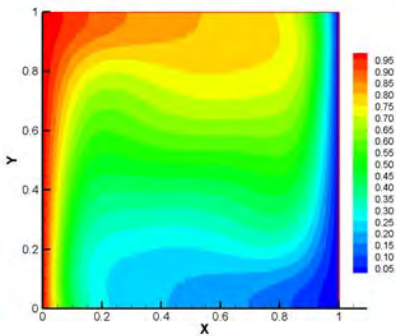

Figure 45 Temperature field in Case 2 [45]

The streamlines temperature field and Nusselt number obtained from the present LBM-MCM approach agree with that of the benchmark solutions well. Thus, the combined LBM-MCM is reliable for the natural convection simulation.

Table 3 Comparison of Nusselt number [45]

| Results | | $Nu_{max}$ | $Y_{Nu_{max}}$ | $Nu_{ave}$ |
|---|---|---|---|---|
| $Ra = 10^4$ | LBM-MCM | 3.50 | 0.15 | 2.21 |
| | Reference [46] | 3.53 | 0.15 | 2.24 |
| | Error | 1.0% | 0 | 1.4% |
| $Ra = 10^5$ | LBM-MCM | 7.38 | 0.08 | 4.38 |
| | Reference [46] | 7.72 | 0.08 | 4.52 |
| | Error | 4.4% | 0 | 3.1% |

## CONCLUSION

In this chapter, meso-macro-multiscale methods are employed to develop different fluid flow and heat transfer problems. Two strategies of multiscale methods exist: for the first one, the whole domain is divided into multiple subdomains and different domains use various numerical methods. Message passing among subdomains decides the accuracy of this type of multiscale numerical method. For the second one, various parameters are solved with different numerical methods.

Two schemes to fulfill LBM-FVM hybrid method are proposed and they are verified by solving lid driven flow. The results indicate that nonequilibrium extrapolation scheme is suitable for the low velocity cases and finite-difference velocity gradient method is valid for the high velocity cases. Based on these results, nonequilibrium extrapolation is employed to combine LBM and FVM for natural convection simulation and various geometric settings are used for the first type multiscale method. The results show that the proposed hybrid method is reliable to solve natural convection problems and geometric settings don't affect the accuracy.

Another strategy to fulfill LBM-FVM multiscale method is proposed to solve convection problems: velocity field is solved using LBM; temperature field is analyzed with FVM. The streamlines and temperature fields obtained by the three methods agree with each other very well for different Rayleigh numbers at $10^4$, $10^5$ and $10^6$. The hybrid method also has a good accuracy for the Nusselt number when comparing with the reference and the two pure methods results. Thus, the hybrid LBM-FVM is reliable for the natural convection simulation. This method is designed to take advantages of both LBM and FVM. It is well known that LBM has its advantage in solving complex geometry fluid flow while FVM has high efficiency for conservative laws. So this hybrid method can show its advantage in the case such as heat transfer in the complex geometry fluid flow problems. Second type of LBM-MCM multiscale method is proposed and verified. The LBM is applied to solve the velocity field and the temperature field is obtained by the MCM. This combined method is employed to solve two cases natural convection in a cavity. The streamlines temperature field and Nusselt number obtained from the present LBM-MCM approach agree with that of the benchmark solutions well. Thus, the combined LBM-MCM is reliable for the natural convection simulation. LBM-FVM and LBM-MCM proposed in this chapter are valid to solve fluid flow and heat transfer problems.

## ACKNOWLEDGEMENT

Support for this work by U.S. National Science Foundation under grant number CBET- 1066917, Chinese National Natural Science Foundations under Grants 51129602 and 51476103, and Innovation Program of Shanghai Municipal Education Commission under Grant 14ZZ134 are gratefully acknowledged.

## NOMENCLATURE

| | |
|---|---|
| $c$ | lattice speed |
| $c_s$ | lattice sound speed |
| $e_i$ | velocity in all directions |
| $f$ | density distribution |
| $F$ | non-dimensional time |
| $\boldsymbol{F}_i$ | body force |
| *Fo* | Fourier number |
| **g** | energy distribution |
| **g** | gravity acceleration $\left(m/s^2\right)$ |
| **G** | effective gravity acceleration $\left(m/s^2\right)$ |
| *H* | height of the enclosure (m) |
| *k* | thermal conductivity (W/m k) |
| $\underline{K}$ | coefficient in Chapman-Enskog expansion |
| $\bar{k}$ | modified thermal conductivity (W/m k) |
| *m* | dimension of the problem |
| *Ma* | Mach number |
| *Nu* | Nusselt number |
| *p* | pressure (N/m²) |
| *P* | non-dimensional pressure |
| *Pr* | Prandtl number |
| **Q** | collision matrix for energy distribution |
| **r** | lattice location vector |
| $R_g$ | Gas constant |
| *Ra* | Rayleigh number |
| *Re* | Reynolds number |
| *RN* | Random number |
| t | time (s) |
| *T* | temperature (K) |
| *u* | horizontal velocity (m/s) |
| $u_0$ | lid velocity (m/s) |
| $u_{sound}$ | sound speed(m/s) |
| *U* | non-dimensional horizontal velocity |
| *v* | vertical velocity (m/s) |
| *V* | non-dimensional vertical velocity |
| **V** | velocity |
| *W* | width of the enclosure (m) |

**Greek Symbols**

| | |
|---|---|
| **a** | lattice acceleration |
| α | thermal diffusivity $\left(m^2/s\right)$ |
| β | volume expansion coefficient of the fluid $\left(K^{-1}\right)$ |
| Γ | coefficient in SIMPLE |
| δ | Kronecker's delta |
| Δ*t* | time step |
| **ε** | lattice particle velocity |
| θ | nondimensional temperature |
| *Π* | moment of order 2 |
| ν | kinematic viscosity (m²/s) |
| μ | dynamic viscosity (N·s/m²) |
| ρ | Density (kg/m³) |
| τ | non-dimensional time |
| $\tau_v$ | relaxation time for velocity |
| $\tau_T$ | relaxation time for energy |

| $\Phi$ | general variable |
| $\omega_i$ | value factor for velocity |
| $\omega_i^T$ | value factor for energy |
| $\Omega$ | collision operator |